\documentclass[fleqn,usenatbib,useAMS]{mnras}

\DeclareRobustCommand{\VAN}[3]{#2}
\let\VANthebibliography\thebibliography
\def\thebibliography{\DeclareRobustCommand{\VAN}[3]{##3}\VANthebibliography}

\usepackage{amsmath,amsfonts,amssymb} 
\usepackage{bm} 
\usepackage[table,  dvipsnames]{xcolor} 
\usepackage{hyperref} 
\usepackage{nicefrac} 
\usepackage{placeins}
\usepackage{microtype}

\usepackage[varg]{txfonts} 
\usepackage[T1]{fontenc}

\usepackage{soul}

\hypersetup{
	pdftitle = {Halting the migration of super-Earths by efficient gap opening in radiative, low viscosity disks},
	pdfauthor = {A. Ziampras, R. P. Nelson, S.-J. Paardekooper},
	pdfsubject = {},
	colorlinks = {true},
	linkcolor = [rgb]{0,0.35,0.7},
	citecolor = [rgb]{0,0.35,0.7},
	filecolor = [rgb]{0.61,0,0},
	urlcolor = [rgb]{0.61,0,0},
}

\usepackage{graphicx}  
\graphicspath{{figures/}}

\newcommand{\tensorGR}[1]{\overline{\bm{{#1}}}}
\newcommand{\DP}[2]{\frac{\partial{#1}}{\partial{#2}}}
\newcommand{\D}[2]{\frac{\text{d}{#1}}{\text{d}{#2}}}

\newcommand{\ap}{a_\mathrm{p}}
\newcommand{\G}{\text{G}}
\newcommand{\Mstar}{M_\star}
\newcommand{\Lstar}{L_\star}
\newcommand{\Rp}{R_\mathrm{p}}
\newcommand{\Mp}{M_\mathrm{p}}
\newcommand{\hp}{h_\mathrm{p}}
\newcommand{\Hp}{H_\mathrm{p}}

\newcommand{\Mth}{M_\mathrm{th}}
\newcommand{\Mf}{M_\mathrm{f}}
\newcommand{\Msun}{\mathrm{M}_\odot}
\newcommand{\Lsun}{\mathrm{L}_\odot}

\newcommand{\Mearth}{\mathrm{M}_\oplus}
\newcommand{\Rgas}{\mathcal{R}}
\newcommand{\cs}{c_\mathrm{s}}

\newcommand{\OmegaK}{\Omega_\mathrm{K}}
\newcommand{\uK}{u_\mathrm{K}}

\newcommand{\taueff}{\tau_\mathrm{eff}}
\newcommand{\kappaR}{\kappa_\mathrm{R}}
\newcommand{\kappaP}{\kappa_\mathrm{P}}

\newcommand{\rhomid}{\rho_\mathrm{mid}}
\newcommand{\sigmaSB}{\sigma_\mathrm{SB}}
\newcommand{\vel}{\bm{u}}
\newcommand{\xh}{{x}_\mathrm{h}}

\newcommand{\Qcool}{Q_\mathrm{cool}}
\newcommand{\Qirr}{Q_\mathrm{irr}}
\newcommand{\Qrad}{Q_\mathrm{rad}}

\newcommand{\Sigmag}{\Sigma_\mathrm{g}}

\newcommand{\St}{\mathrm{St}}

\newcommand{\Rh}{R_\mathrm{H}}
\newcommand{\xshock}{x_\mathrm{sh}}

\newcommand{\pluto}{\texttt{PLUTO}}
\newcommand{\fargo}{{\texttt{FARGO3D}}}

\defcitealias{MCNALLY-ETAL-2019A}{MN19a}

\title[Planet migration in nonlinear regimes]{Halting the migration of super-Earths by efficient gap opening in radiative, low viscosity disks}

\author[A.~Ziampras et al.]{Alexandros~Ziampras$^{1,2,3}$, 
Richard~P.~Nelson$^{2}$\thanks{E-mail: r.p.nelson@qmul.ac.uk},
Sijme-Jan~Paardekooper$^{4}$
\\
$^{1}$Ludwig-Maximilians-Universit{\"a}t M{\"u}nchen, Universit{\"a}ts-Sternwarte, Scheinerstr.~1, 81679 M{\"u}nchen, Germany\\
$^{2}$Astronomy Unit, Dept. of Physics and Astronomy, Queen Mary University of London, London E1 4NS, UK\\
$^{3}$Max Planck Institute for Astronomy, K{\"o}nigstuhl 17, 69117 Heidelberg, Germany\\
$^{4}$TU Delft, Faculty of Aerospace Engineering, Kluyverweg 1, 2629 HS Delft, The Netherlands
}

\date{Accepted XXX. Received YYY; in original form ZZZ}
\pubyear{2023}

\begin{document}
\label{firstpage}
\pagerange{\pageref{firstpage}--\pageref{lastpage}}
\maketitle

\begin{abstract}
	While planet migration has been extensively studied for classical viscous disks, planet--disk interaction in nearly inviscid disks has mostly been explored with greatly simplified thermodynamics. In such environments, motivated by models of wind-driven accretion disks, even Earth-mass planets located interior to 1\,au can significantly perturb the disk, carving gaps and exciting vortices on their edges. Both processes are influenced by radiative transfer, which can both drive baroclinic forcing and influence gap opening. We perform a set of high-resolution radiation hydrodynamics simulations of planet--disk interaction in the feedback and gap-opening regimes, aiming to understand the role of radiation transport in the migration of super-Earth-mass planets representative of the observed exoplanet population. We find that radiative cooling drives baroclinic forcing during multiple stages of the planet's migration in the feedback regime ($\sim1.5\,\Mearth$), significantly delaying the onset of vortex formation at the gap edge but ultimately resulting in type-III runaway migration episodes. For super-thermal-mass planets ($\sim 6.7\,\Mearth$), radiative cooling is fundamentally linked to the gap opening process, with the planet stalling instead of undergoing vortex-assisted migration as expected from isothermal or adiabatic models.
    This stalling of migration can only be captured when treating radiative effects, and since it affects super-thermal-mass planets its implications for both the final configuration of planetary systems and population synthesis modeling are potentially huge.
    Combining our findings with previous related studies, we present a map of migration regimes for radiative, nearly-inviscid disks, with the cooling-mediated gap-opening regime playing a central role in determining the planet's orbital properties.
\end{abstract}

\begin{keywords}
    planet--disc interactions --- accretion discs --- hydrodynamics --- radiation: dynamics --- methods: numerical
\end{keywords}


\section{Introduction}
\label{sec:introduction}

The direct observation of two nascent exoplanets in the protoplanetary disk around PDS~70 \citep{keppler-etal-2018,haffert-etal-2019} has cemented the idea that planets are born in circumstellar disks. With over 5700 confirmed exoplanets as of today\footnote{\url{https://exoplanetarchive.ipac.caltech.edu/}}, the community now has enough data to test different formation theories in order to interpret the orbital parameters, composition, and multiplicity of observed exoplanets.

One aspect critical to understanding planet formation is the dynamical interaction between young planets and their surrounding gaseous disk in a process termed planet migration \citep[e.g.,][]{goldreich-tremaine-1979,lin-papaloizou-1993,tanaka-etal-2002}. Planets interact gravitationally with the disk, exciting spiral arms \citep{ogilvie-lubow-2002} throughout the disk as well as corotating horseshoe flows in the vicinity the planet, which in turn exert torques that can induce inward or outward migration \citep{goldreich-tremaine-1980,kley-nelson-2012}. For a recent review, we refer the reader to \citet{paardekooper-etal-2022}.

As planet-driven spiral arms propagate with a pattern speed that matches the orbital frequency of the planet, they eventually become transsonic with respect to the background flow, steepening into shocks. Their dissipation then drives a local angular momentum flux \citep{goodman-rafikov-2001} that can lead to the carving of a gap around the planet's orbit, weakening the aforementioned torques and substantially slowing down migration \citep{rafikov-2002}. This effect is particularly strong in disks that have low viscosity, where refilling of the gap through viscous diffusion is not efficient. With that in mind, different regimes of planet migration have been identified.

Following \citet{rafikov-2002b}, \citet{MCNALLY-ETAL-2019A} (hereafter \citetalias{MCNALLY-ETAL-2019A}) defined three regimes in low viscosity disks based on the planet mass $\Mp$ relative to the thermal mass $\Mth$ and the feedback mass $\Mf$, with
\begin{equation}
    \label{eq:thermal-mass}
    \Mth = \frac{2h^3}{3}\Mstar, \quad\Mf \approx 3.8 \left(\frac{Q}{h}\right)^{-5/13} \Mth, \qquad Q = \frac{\cs\OmegaK}{\pi G\Sigma}.
\end{equation}
Here, $Q$ is the Toomre parameter \citep{toomre-1964}, with the remaining symbols introduced in Sect.~\ref{sec:physics-numerics}. The thermal mass indicates the planet mass above which the wakes shock straight from launch, and the feedback mass indicates a minimum mass for the planet to significantly modify the gas density in its surroundings. For thin disks ($h\ll1$) stable against gravitational instabilities ($Q\gg1$), we have $\Mf<\Mth$. The three regimes are then:
\begin{itemize}
    \item[] \textbf{Type-I}: for planets with mass $\Mp<\Mf$, planet--disk interaction is limited to the excitation of (Lindblad) spiral arms and the formation of a corotating region around the planet, with both features exerting a torque on the planet \citep{goldreich-tremaine-1980}. Given the low-level perturbations in this regime, torque prescriptions for both Lindblad and corotation torques exist \citep{paardekooper-etal-2011} and can be readily used in population synthesis models \citep[e.g.][]{coleman2014,izidoro2021,emsenhuber2021}.
    \item[] \textbf{Feedback}: for planets with $\Mf<\Mp<\Mth$, the dissipation of spiral arms can alter the surface density profile near the planet's orbit, carving a shallow gap and slowing down migration. In that sense, the influence of the planet on the disk structure can ``feed back'' on the migration of the planet, which in turn can influence gap opening. The idea that gap opening can feed back onto planet migration and cause it to stall when the planet reaches the ``inertial limit'' goes back to \cite{hourigan-ward-1984} and \cite{ward-hourigan-1989}. In this context, the planet forms an asymmetric gap structure that is deeper exterior to the planet's orbit (see also Fig.~\ref{fig:lowmass-adb-gap-structure}) and results in the negative outer Lindblad torque being progressively weakened, until the planet theoretically stalls. However, \citetalias{MCNALLY-ETAL-2019A} have shown that, for simplified thermodynamics, this modified surface density profile is Rossby-wave unstable \citep{lovelace-1999}. This leads to the growth of numerous small-scale vortices that act to diffusively smooth gap edges, effectively assisting and sustaining inward migration, with episodes of type-III runaway migration also being observed.
    \item[] \textbf{Type-II}: for more massive planets ($\Mp>\Mth$), the spiral wakes shock as they are launched, resulting in deep gap opening for sufficiently low turbulent diffusivity \citep{crida-etal-2006} that can also be subject to various instabilities \citep[e.g.,][]{kanagawa-etal-2015,hallam-paardekooper-2017,muley-etal-2024}. In this regime the planet is expected to practically stall, trending inwards on the viscous accretion timescale of the disk if the latter has any viscosity \citep{lin-papaloizou-1986,ward-1997b} or at a rate set by the diffusivity at the gap edge \citep[e.g.,][]{MCNALLY-ETAL-2019A,lega-etal-2021}. This stalling has been observed in recent simulations of very low viscosity disks containing Jovian mass planets orbiting at 5~au \citep{lega-etal-2021,lega-etal-2022}.
\end{itemize}
For $\Mstar=1\,\Msun$ and typical disk parameters used in the literature ($h\sim0.05$, $Q\sim100$), the feedback and thermal masses amount to $\Mf\sim6\,\Mearth$ and $\Mth\sim30\,\Mearth$, respectively. With the median planet mass from the Kepler and K2 missions being $\sim 5.6\,\Mearth$\footnote{\url{https://exoplanetarchive.ipac.caltech.edu/}}, this calculation points to the type-I regime as the most relevant for the ``typical'' super-Earth. 

\begin{figure}
    \includegraphics[width=\columnwidth]{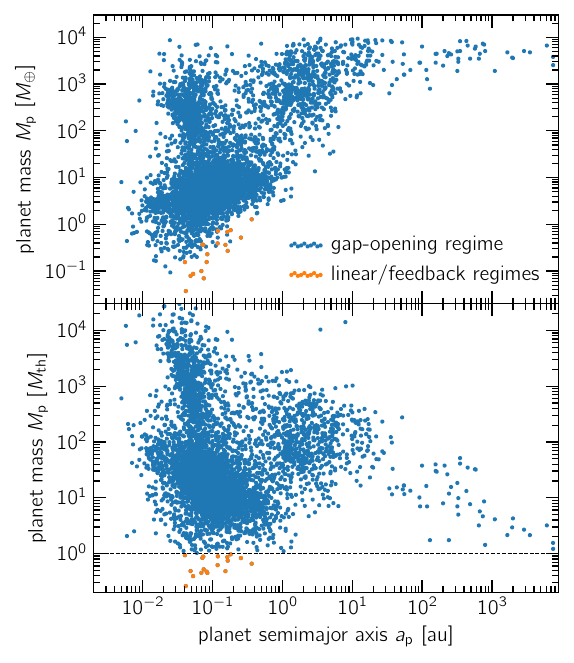}
    \caption{Top: confirmed exoplanets as of July 2025 from the NASA exoplanet archive. Bottom: the same population in the context of the thermal mass $\Mth$ in Eq.~\eqref{eq:thermal-mass}, computed assuming the stellar parameters given by the exoplanet data and a passive, irradiated disk model ($\alpha=0$) following Sect.~\ref{sec:physics-numerics}. The vast majority of planets are above the thermal mass. We note that this is a simplistic calculation that ignores the structure of the disk inner rim ($\lesssim0.1$\,au). We also note that the actual measurements during planetary transits (e.g., with the Kepler telescope) are planetary radius and orbital period rather than mass and semimajor axis. The latter are inferred using the model of \citet{chen-kipping-2017} and Keplerian rotation, respectively, when a direct mass measurement is not available.}
    \label{fig:population}
\end{figure}

However, the aforementioned disk conditions are only applicable in the context of traditionally turbulent, ``viscously'' heated disks, where magnetohydrodynamical (MHD) processes such as the magnetorotational instability \citep[MRI,][]{balbus-hawley-1991} can drive significant turbulent heating through dissipation and accretion through radial angular momentum transport. As recent work has shown, non-ideal MHD effects can quench or significantly suppress the MRI in the bulk of the disk, leading to a laminar, practically inviscid ``dead zone'' \citep{gammie-1996,bai-stone-2013,gressel2015}. This leaves stellar irradiation as the dominant heat source in the disk, and results in drastically cooler disks with $h\sim0.02$--0.03 in the 1--5\,au range \citep{chiang-goldreich-1997}. The feedback and thermal masses are then reduced to $\Mf\sim0.5$--$1.5\,\Mearth$ and $\Mth\sim2$--6$\,\Mearth$, respectively. This implies that a significant fraction of super-Earths must have at the very least experienced feedback effects, and possibly opened deep gaps in their disks, between their formation and the end of the disk lifetime (see Fig.~\ref{fig:population}).

It is worth noting that the low-viscosity nature of protoplanetary disks has two important effects: not only does the lack of dissipation lead more readily to disordered flow induced by the planet, i.e. vortices, it also changes the underlying background flow, and in particular the temperature profile. Together, they conspire to make planet migration very different in low-viscosity disks compared to traditional viscous accretion disks.

From the above, it becomes clear that the feedback and gap-opening regimes are of particular interest for the study of super-Earth migration in low-turbulence, passively heated disks. While \citetalias{MCNALLY-ETAL-2019A} have provided a detailed analysis of the three regimes in both viscous and inviscid disks, their work is based on a simplified thermodynamical model that does not account for the effects of radiative transfer. In this work, we build on their results by performing a set of high-resolution radiation hydrodynamical simulations of super-Earth migration in the feedback and gap-opening regimes for realistic disk conditions, and highlight the importance of radiative processes in deciding the fate of migrating planets.

We outline our physical framework and numerical setups in Sect.~\ref{sec:physics-numerics}. In Sect.~\ref{sec:comparison-to-mcnally} we establish a connection to the findings of \citetalias{MCNALLY-ETAL-2019A}, and present the results of our models in Sects.~\ref{sec:feedback-regime}~\&~\ref{sec:gap-opening-regime}. We present a map of migration regimes for nearly inviscid, radiative disks in Sect.~\ref{sec:map}, and discuss its applications and limits. We then discuss our findings in Sect.~\ref{sec:discussion}, and conclude with a summary in Sect.~\ref{sec:summary}.

\section{Physics and numerics}
\label{sec:physics-numerics}

In this section we outline our physical framework and describe our numerical setup. While the physical model is largely based on the one presented in \citet{ziampras-etal-2024a}, we provide a brief summary of the key equations and assumptions for completeness.

\subsection{Physical model}
\label{sub:physics}

We consider a vertically integrated disk of perfect gas with adiabatic index $\gamma=7/5$ and mean molecular weight $\mu=2.353$ around a star with mass $\Mstar=1\,\Msun$ and luminosity $\Lstar$. The gas has surface density $\Sigma$, velocity field $\vel$ and pressure $P=(\gamma-1)e$, with $e$ being the vertically integrated thermal energy density. In this framework, the Navier--Stokes equations read
\begin{subequations}
\label{eq:navier-stokes}
\begin{align}
	\label{eq:navier-stokes-1}
	\D{\Sigma}{t} =-\Sigma\nabla\cdot\vel,
\end{align}
\begin{align}
\label{eq:navier-stokes-2}
	\D{\vel}{t} =-\frac{1}{\Sigma}\nabla P -\nabla(\Phi_\star+\Phi_\mathrm{p}) + \nabla\cdot\tensorGR{\sigma},
\end{align}
\begin{align}
\label{eq:navier-stokes-3}
	\D{e}{t} =-\gamma e\nabla\cdot\vel + Q.
\end{align}
\end{subequations}
In the above, $\Phi_\star=-\G \Mstar/R$ is the gravitational potential of the star at radial distance $R$, $\Phi_\mathrm{p}$ is the potential of an embedded planet with mass $\Mp$ and semi-major axis $\ap$, and $\tensorGR{\sigma}$ is the viscous stress tensor. The isothermal sound speed is given by $\cs=\sqrt{P/\Sigma}$ and relates to the temperature as $T=\mu\cs^2/\Rgas$. We can then define the pressure scale height $H=\cs/\OmegaK$ and the aspect ratio $h=H/R$, where $\OmegaK=\sqrt{\G\Mstar/R^3}$ is the Keplerian orbital frequency. The gravitational constant and gas constant are denoted by $\G$ and $\Rgas$, respectively.

The source term $Q$ in the energy equation encapsulates viscous dissipation, stellar irradiation, surface cooling and in-plane radiative diffusion with:
\begin{subequations}
    \label{eq:source-terms}
    \begin{align}
        \label{eq:source-terms-1}
        Q_\mathrm{visc} = \frac{1}{2\nu\Sigmag}\mathrm{Tr}(\tensorGR{\sigma}^2) \approx \frac{9}{4}\nu\Sigmag\OmegaK^2,\quad \nu=\alpha\sqrt{\gamma}\cs H,
    \end{align}
    \begin{align}
        \label{eq:source-terms-2}
        Q_\mathrm{irr} = 2\frac{\Lstar}{4\pi R^2} (1-\epsilon)\frac{\theta}{\taueff}, \quad \theta = R\D{(\chi h)}{R}\approx \frac{2\chi h}{7},
    \end{align}
    \begin{align}
        \label{eq:source-terms-3}
        Q_\mathrm{cool} = -2\frac{\sigmaSB T^4}{\taueff}, \quad \taueff = \frac{3\tau}{8} + \frac{\sqrt{3}}{4} + \frac{1}{4\tau}, \quad \tau = \frac{1}{2}\kappa\Sigma,
    \end{align}
    \begin{align}
        \label{eq:source-terms-4}
        Q_\mathrm{rad} = \sqrt{2\pi}H\nabla\cdot \left(\lambda\frac{4\sigmaSB}{\kappa\rhomid}\nabla T^4\right), \quad \rhomid = \frac{1}{\sqrt{2\pi}}\frac{\Sigma}{H}.
    \end{align}
\end{subequations}
Here, we have adopted the $\alpha$-viscosity prescription of \citet{shakura-sunyaev-1973} with $\alpha=10^{-6}$, the irradiation model of \citet{menou-goodman-2004} with a disk albedo $\epsilon=1/2$, an effective optical depth $\taueff$ following \citet{hubeny-1990} that depends on the Rosseland and Planck mean opacities $\kappaR=\kappaP=\kappa$ \citep{lin-papaloizou-1985}, and the flux-limited diffusion (FLD) approximation of \citet{levermore-pomraning-1981} for radiative diffusion with the flux limiter $\lambda$ by \citet{kley-1989}. Following \citet{chiang-goldreich-1997}, we set the height of the irradiation surface $\chi=z_\text{irr}/H = 4$. The Stefan--Boltzmann constant is denoted with $\sigmaSB$. We refer the reader to \citet{ziampras-etal-2023a} for a more detailed description on the radiative terms considered here.

The planet is treated as a Plummer potential centered at $\bm{R}_\text{p}$ with
\begin{equation}
	\label{eq:plummer-potential}
	\Phi_\mathrm{p} = -\frac{\G\Mp}{\sqrt{d^2+\epsilon^2}},\qquad \bm{d}=\bm{R}-\bm{R}_\mathrm{p},
\end{equation}
and $\epsilon=0.6\Hp$ is the softening length, accounting for the vertical stratification of the disk \citep{mueller-kley-2012}. The planet is initialized with a semimajor axis of 1\,au and on a circular orbit. To work in a star-centered frame, we include the indirect terms due to the star--planet system orbiting its center of mass as well as due to the star--disk interaction. Since we do not consider the effects of self-gravity on the disk, we subtract the azimuthally averaged surface density before computing the gravitational torque by the disk on the planet \citep{baruteau-masset-2008b}.

A useful quantity in the context of planet migration in the feedback and gap-opening regimes is the potential vorticity or vortensity $\varpi$, defined as
\begin{equation}
    \label{eq:vortensity}
    \varpi = \frac{\nabla\times\vel}{\Sigma}.
\end{equation}
For an unperturbed, Keplerian disk, the Keplerian vortensity is then $\varpi_\text{K} = 0.5\,\OmegaK/\Sigma$. We will use these quantities to track the activity of vortices, which are of key importance in the regimes studied here as they provide the radial turbulent mixing necessary to assist inward migration, at least in the feedback regime \citepalias{MCNALLY-ETAL-2019A}. In the gap-opening regime, $\varpi$ can be used both as a tracer of vortex formation at the Rossby-wave unstable gap edge, as well as a proxy for the inverse surface density $\Sigma^{-1}\propto \varpi$ once a clear gap profile has been established.

\subsection{Numerical setup}
\label{sub:numerics}

We solve the equations described in Sect.~\ref{sub:physics} using the Godunov code \pluto{} \citep{mignone-etal-2007}. The term $\Qrad$ in Eq.~\eqref{eq:source-terms-4} is implemented following \citet{ziampras-etal-2020a}, and the orbital evolution of the planet with the N-body module detailed in \citet{thun-kley-2018}. We further use the FARGO algorithm \citep{masset-2000,mignone-etal-2012}, which relaxes the strict timestep constraint due to the rapidly rotating inner boundary while also improving numerical accuracy by subtracting the background Keplerian flow before solving for advection. Viscous diffusion and dissipation are implemented with a super-time-stepping scheme \citep[STS,][]{alexiades-etal-1996}.

Finally, we use the HLLC Riemann solver \citep{toro-etal-1994}, the flux limiter by \citet{vanleer-1974}, a third-order weighted essentially non-oscillatory (WENO) reconstruction method \citep{yamaleev-carpenter-2009}, and a second-order Runge--Kutta time-stepping scheme, combined with the \texttt{CHAR\_LIMITING} option for improved accuracy. This combination of numerical choices has been shown to match results obtained with the code \fargo{} \citep{benitez-llambay-etal-2016} regarding planetary migration tracks \citep{afkanpour-etal-2024} as well as the evolution of vortensity near the planet's corotating region \citep{ziampras-etal-2023b} in the low-viscosity regime.

We utilize a cylindrical polar grid for all simulations with $R\in[0.3,2]$\,au and $\phi\in[0,2\pi]$, with logarithmic spacing in the radial direction. Wave-damping zones are applied for $R<0.39$\,au and $R>1.53$\,au \citep{benitez-etal-2016} following the prescription of \citet{devalborro-etal-2006} with a damping timescale $t_\mathrm{damp}=0.1\,\OmegaK^{-1}$. At the radial boundary edges all quantities are reset to their initial values. We maintain a resolution of approximately 25 cells per scale height in both the radial and azimuthal directions at $R=1$\,au, which translates to $N_R\times N_\phi=1296\times4320$ cells for the models in Sect.~\ref{sec:comparison-to-mcnally} and $1944\times6480$ cells for all other models.

Our initial conditions follow power-laws in surface density and temperature with
\begin{equation}
    \label{eq:initial-conditions}
    \Sigma_0 = \Sigma_\text{ref}\left(\frac{R}{1\,\text{au}}\right)^s,\quad T_0 = T_\text{ref}\left(\frac{R}{1\,\text{au}}\right)^q \Rightarrow h_0 = h_\text{ref}\left(\frac{R}{1\,\text{au}}\right)^\frac{q+1}{2},
\end{equation}
with the gas velocity initialized as sub-Keplerian, accounting for the radial pressure gradient:
\begin{equation}
    \label{eq:initial-velocity}
    u_{R,0} = 0, \qquad u_{\phi,0} = \OmegaK R\sqrt{1+(s+q)h_0^2}.
\end{equation}
The planet is introduced at $t=0$, growing to its final mass over 10~orbits using the formula in \citet{devalborro-etal-2006}. Since $\Mp$, $\Sigma_\text{ref}$, and $h_\text{ref}$, as well as the power-law indices $s$ and $q$ are problem-dependent, we will mention their values in their respective sections.

\section{Comparison to \citet{mcnally-etal-2019a}}
\label{sec:comparison-to-mcnally}

In their study of planet migration in the feedback regime, \citetalias{MCNALLY-ETAL-2019A} highlighted the sensitivity of the results to the numerical resolution, showing that the planet's migration track does not converge even qualitatively for a resolution of 93 cells per scale height in inviscid setups. This finding underscores the problems associated with numerical diffusion, and necessitates incorporating at the very least a floor value of physical diffusivity such that numerical experiments are well-posed. In fact, they showed that for a value of $\nu=10^{-9}\,\text{au}^2\OmegaK^\text{1\,au}$, or $\alpha \approx 10^{-6}$, they obtain similar results to their inviscid models, while for a value of $\alpha\sim10^{-4}$ their results show excellent convergence even at their lowest resolution of 23 cells per scale height. This suggests that a floor value of $\alpha=10^{-6}$ is necessary to shield our models from the effects of numerical diffusion.


Nevertheless, in the interest of examining the extent to which \pluto{} can reproduce the features observed in \citetalias{MCNALLY-ETAL-2019A}, we mirror their fiducial setup of 23 cells per scale height in a globally isothermal, barotropic, inviscid disk. We therefore choose, for this comparison simulation only, $\Mp=1.25\times10^{-5}\,\Msun=4.2\,\Mearth\approx0.44\,\Mth\approx2\,\Mf$, $\alpha=0$, $s=-1.5$, $q=0$, $\Sigma_\text{ref} = 1700\,\text{g}/\text{cm}^2$, and $h_\text{ref} = 0.035$ (i.e., $T_\text{ref}=307.5$\,K, see Sect.~\ref{sub:numerics}). For this run alone, we also set the smoothing length to $\epsilon=0.4\Hp$ in Eq.~\eqref{eq:plummer-potential}.

\begin{figure}
    \includegraphics[width=\columnwidth]{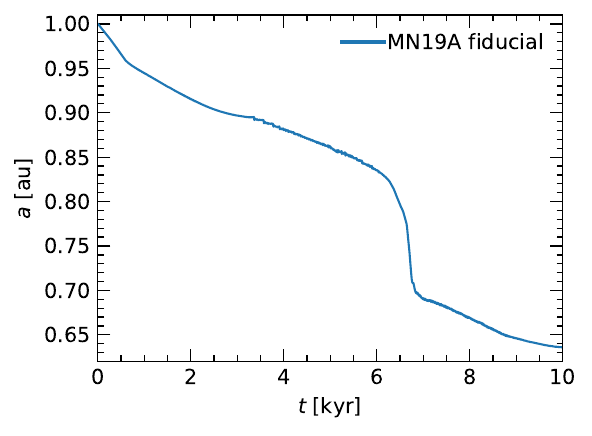}
    \caption{Migration track of a planet in an isothermal model aiming to reproduce the fiducial setup of \citetalias{MCNALLY-ETAL-2019A} (see ``run r2'' in Fig.~2 therein) with excellent agreement. The planet slows down to a halt by $t\sim 3$\,kyr, before transitioning to vortex-assisted migration and an eventual runaway episode at $t\sim6.5$\,kyr.}
    \label{fig:migration-colin}
\end{figure}

The planet's migration track after 10\,kyr is shown in Fig.~\ref{fig:migration-colin}, showing remarkable agreement both qualitatively and quantitatively with the fiducial model of \citetalias{MCNALLY-ETAL-2019A} (see ``run r2'' in Fig.~2 therein). In particular, we recover with excellent accuracy:
\begin{itemize}
    \item the transition from type-I to the slower, feedback-mediated migration rate at $t\sim0.4$\,kyr,
    \item the complete halting of the planet at $t\sim3$\,kyr before transitioning to vortex-assisted migration,
    \item the type-III episode at $t\sim6.5$\,kyr and the immediate resumption of vortex-assisted migration thereafter, and
    \item the overall migration distance of $\sim0.33$\,au after 8\,kyr.
\end{itemize}
Considering that the numerical methods used in \pluto{} are fundamentally different from those of \fargo{}, which \citetalias{MCNALLY-ETAL-2019A} used, we conclude that our setup is suitable to build on their work and that our numerical experiments will yield qualitatively sound results after adopting a reasonable floor value of $\alpha$.

\section{The feedback regime: $\Mf<\Mp<\Mth$}
\label{sec:feedback-regime}

Now that a connection to the results of \citetalias{MCNALLY-ETAL-2019A} has been established in Sect.~\ref{sec:comparison-to-mcnally}, we re-examine planet migration in the feedback mass regime but while taking into account more realistic disk conditions and radiative processes (see Eq.~\eqref{eq:source-terms}). To do this, we need to make some adjustments to both our methods and initial conditions.

For one, we are interested in the inner few au of the disk, a region which is typically optically thick. This makes the disk radiatively inefficient \citep{bae-etal-2021, ziampras-etal-2024b}, such that the isothermal assumption is invalid. In order to establish a benchmark to compare our radiative simulations against, an adiabatic model is more suitable instead. We therefore choose to also carry out an adiabatic simulation where we evolve Eq.~\eqref{eq:navier-stokes-3} with $Q=0$ and by further enabling the flag \texttt{ENTROPY\_SWITCH} in \pluto{} to prevent the accumulation of heat due to shock heating \citep{rafikov-2016}.

Secondly, since we aim to model a passively irradiated disk in its class-II stage while still loosely maintaining a qualitative connection to \citetalias{MCNALLY-ETAL-2019A}, we maintain the MMSN-like $\Sigma_\text{ref}=1700\,\text{g}/\text{cm}^2$ but choose $s=-1$. For the temperature profile, we assume a balance between stellar irradiation and surface cooling for a star with the luminosity of the Sun at an age of 1\,Myr following the Hayashi track \citep[$\Lstar=1.78\,\Lsun$,][]{hayashi-1981}. In equilibrium ($\Qirr=\Qcool$), this yields $q=-3/7$ and $h_\text{ref}=0.025$, or $T_\text{ref}\approx157\,\text{K}$. As discussed in Sect.~\ref{sec:comparison-to-mcnally}, we use $\alpha=10^{-6}$ to mitigate the effects of numerical diffusion. We nevertheless carry out a higher-resolution model in Appendix~\ref{apdx:resolution} to verify that our results are robust against numerical diffusion.

Since we are using a lower value for $h_\text{ref}$, we adjust the planet's mass to $\Mp=4.6\times10^{-6}\,\Mstar = 1.5\,\Mearth$, which still corresponds to $\Mp\approx0.44\,\Mth\approx2\,\Mf$. This allows us to compare our results qualitatively to the model in Sect.~\ref{sec:comparison-to-mcnally}. It is also worth stressing that such a low planet mass amounts to only about a quarter of the median mass of the observed super-Earths, highlighting the importance of modeling the planet--disk interaction process with realistic hydrodynamics.

It should be noted that for our choice of parameters the planet can be subject to a dynamical corotation torque due to the radial vortensity gradient \citep{ward-1991,paardekooper-papaloizou-2009}, in addition to baroclinic effects due to the radial temperature gradient \citep[e.g.,][]{pierens-2015,ziampras-etal-2024a}, both of which are absent in the model in Sect.~\ref{sec:comparison-to-mcnally}. Nevertheless, our choices remain motivated as we are interested in the fate of protoplanets migrating in realistic disk conditions, where both of these conditions are met. For a series of more controlled numerical experiments, albeit with simplified thermodynamics, we refer the reader to \citetalias{MCNALLY-ETAL-2019A}.

\begin{figure}
    \includegraphics[width=\columnwidth]{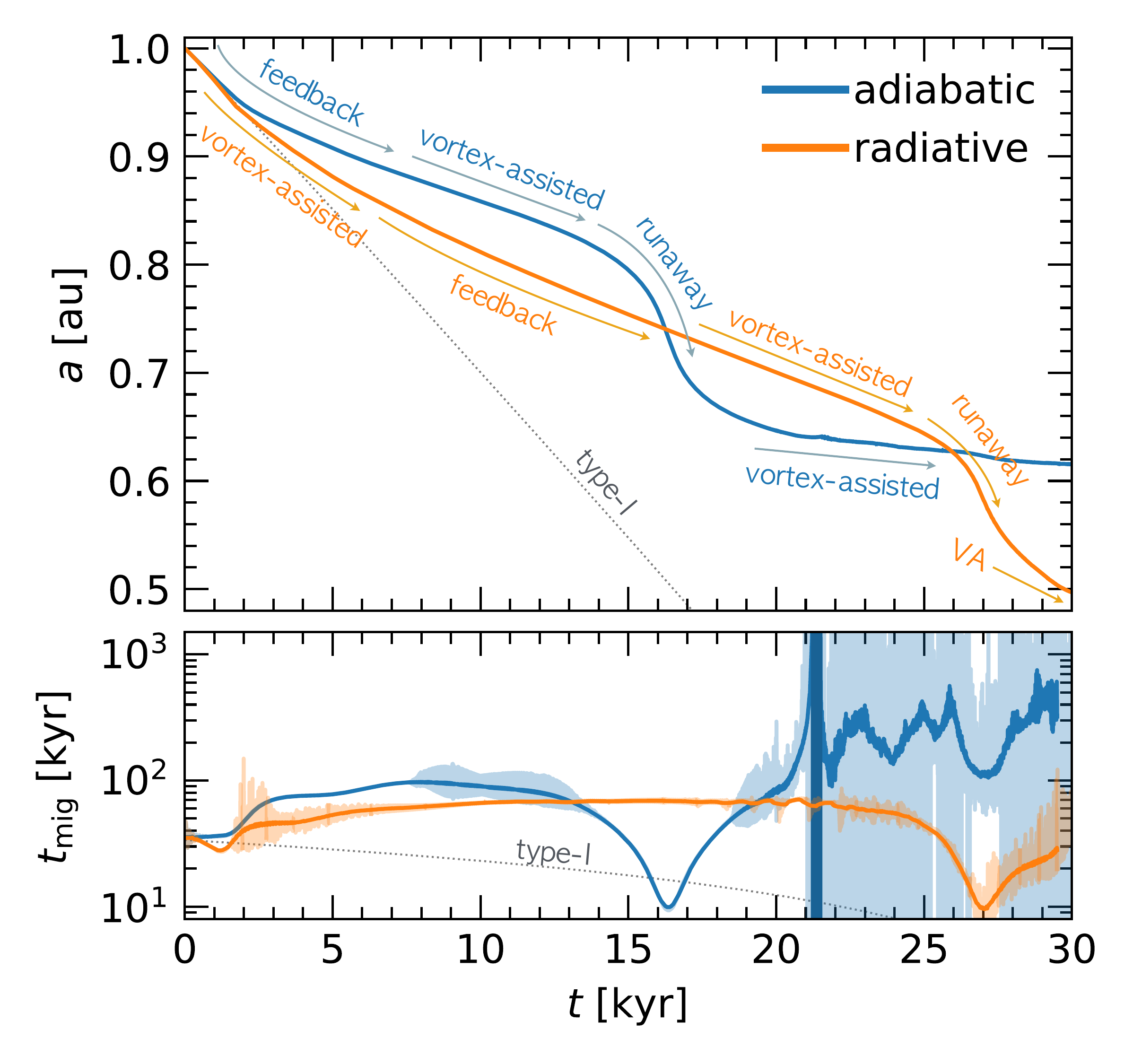}
    \caption{Top: migration tracks for our adiabatic (blue) and radiative (orange) models, with annotations highlighting the different phases of migration. Both runs, while quantitatively different, exhibit periods of vortex-assisted migration and type-III (runaway) episodes. Bottom: the migration timescale $t_\text{mig}$ for both models, smoothed with a 200-year rolling average. The raw data is shown in pale colors, hinting at turbulence due to vortex activity.}
    \label{fig:migration-feedback}
\end{figure}

We first present the migration tracks for the two models in this regime in Fig.~\ref{fig:migration-feedback}, along with the migration timescale computed as $t_\text{mig}=\ap/\dot{a}_\text{p}$. Two takeaways are immediately visible from this figure.

On one hand, both models show qualitatively similar behavior overall when compared to the simulation presented in Sect.~\ref{sec:comparison-to-mcnally}, in that the same features are observed:
\begin{itemize}
    \item the planets initially migrate at the type-I rate before slowing down due to feedback effects,
    \item a steady inward migration track is maintained due to the presence of small-scale vortices, and
    \item both planets eventually experience a runaway inward migration episode at $t\sim16$\,kyr and 27\,kyr for the adiabatic and radiative runs, respectively.
\end{itemize}

On the other hand, the planet's migration tracks are quantitatively different between the two models. While the adiabatic model resembles the isothermal run in Sect.~\ref{sec:comparison-to-mcnally} more closely, the radiative model exhibits:
\begin{itemize}
    \item an initial delay before transitioning to the slower, feedback-mediated regime,
    \item a significantly longer phase of steady inward migration at approximately the same rate as the adiabatic run, and
    \item a turbulent phase of fast inward migration immediately after the runaway episode, discernible by the noise in the bottom panel of Fig.~\ref{fig:migration-feedback} at $t\gtrsim27$\,kyr.
\end{itemize}

In the following paragraphs, we analyze the similarities and differences between the two models in more detail. The analysis is separated into three segments covering:
\begin{enumerate}
    \item the progression into the feedback regime for both models,
    \item the transition from the turbulent vortex-assisted phase to a runaway migration episode for the adiabatic run,
    \item the same development as (ii) for the radiative run.
\end{enumerate}

\subsection{Progression from type-I to feedback regime}
\label{sub:feedback-regime}

After a brief period of migration at the type-I rate up to $t\approx1.5$\,kyr, the planet's feedback onto the disk becomes noticeable in the form of a shallow gap and therefore slower migration. This is more easily discernible in the adiabatic model, where the planet's migration rate is reduced by a factor of $\sim2$--2.5 compared to the type-I rate within the first 5\,kyr. 

\begin{figure}
    \includegraphics[width=\columnwidth]{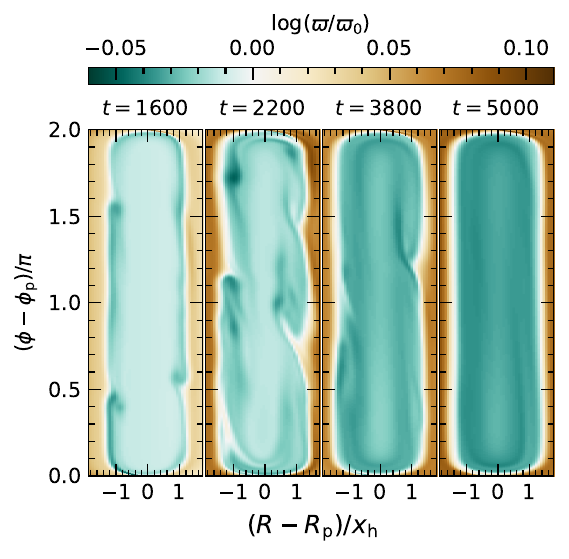}
    \caption{Perturbed vortensity heatmaps in the planet's horseshoe region for the radiative model at different snapshots. Baroclinic forcing due to radiative cooling within the horseshoe region leads to the formation of vortices at $t\sim1.6$\,kyr, which temporarily accelerate the planet but dissipate by $t\sim5$\,kyr.}
    \label{fig:lowmass-rad-baroclinic}
\end{figure}

In the radiative model a gradual slowdown is also visible over 15\,kyr, although the planet migrates at a speed closer to the type-I rate for approximately 4\,kyr instead. This happens due to baroclinic forcing during the formation of the horseshoe region around the planet \citep[see also][]{ziampras-etal-2024a}, which results in a set of small-scale vortices spawning at the separatrix between the corotating region and the background disk (see Fig.~\ref{fig:lowmass-rad-baroclinic}). This interface is located at $R=\Rp\pm \xh$, where $\xh$ is the half-width of the horseshoe region \citep{paardekooper-etal-2010}:
\begin{equation}
    \label{eq:horseshoe-width}
    \xh = \frac{1.1}{\gamma^{1/4}}\left(\frac{0.4}{\epsilon/H}\right)^{1/4} \sqrt{\frac{\Mp}{h\Mstar}}\Rp\Rightarrow \xh\sim 0.013\,\Rp.
\end{equation}
These vortices diffuse the material around the planet's orbit, slowing down the gap opening process, and allowing the planet to sustain the type-I migration rate for a longer period of time. This mechanism cannot be sustained once the horseshoe region is fully established, and the planet's migration rate slows down over time as expected in the feedback regime.

\begin{figure}
    \includegraphics[width=\columnwidth]{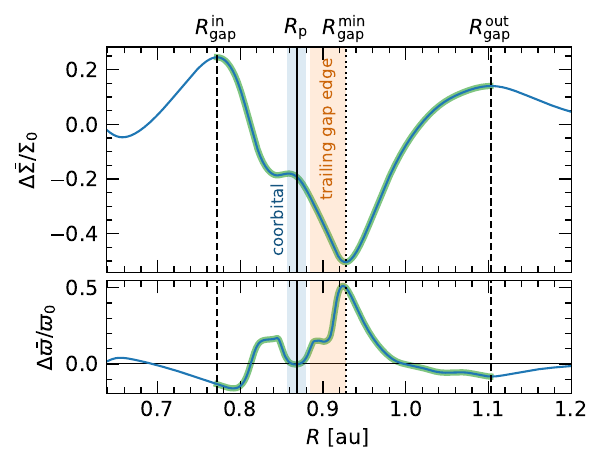}
    \caption{Asymmetric gap structure in the adiabatic model at $t=9$\,kyr showing the azimuthally averaged perturbed surface density (top) and vortensity (bottom). The planet (at $R=\Rp$) is found closer to the inner gap edge, with a pile-up of gas ahead of it and a deep cavity behind it. Blue and orange bands mark the coorbital region $(R\in [\Rp-x_\text{h},\Rp+x_\text{h}])$ and the trailing gap edge $(R\in[\Rp+x_\text{sh},R_\text{gap}^\text{min}])$, respectively, following Eqs.~\eqref{eq:horseshoe-width} and \eqref{eq:shock-width}. The gap region $R\in[R_\text{gap}^\text{in},R_\text{gap}^\text{out}]$ is highlighted with green lines.}
    \label{fig:lowmass-adb-gap-structure}
\end{figure}

As the planet--disk interaction process continues, an asymmetric gap is established, with the planet sitting near the inner gap edge (see Fig.~\ref{fig:lowmass-adb-gap-structure}). This behavior is a consequence of the process that would lead to the ``inertial limit'' discussed by \citet{hourigan-ward-1984} and \citet{ward-hourigan-1989}. The relatively low planet mass with respect to the thermal mass results in a noticeable radial drift of the planet relative to the center of the (shallow) gap, and the gap opening process forms a pile-up of gas ahead of the planet and a deep cavity behind it. The combination of these two effects disproportionately enhances the (positive) inner Lindblad torque, slowing down the planet's migration rate.

Following Fig.~\ref{fig:lowmass-adb-gap-structure} we define several regions of interest that will be helpful in understanding the migration patterns seen in either model. The inner and outer \emph{gap edges} $R_\text{gap}^\text{in}$ and $R_\text{gap}^\text{out}$ are defined as the radial locations where the azimuthally averaged perturbed surface density $\Delta\bar{\Sigma}/\Sigma_0$ is highest in the inner ($R<\Rp$) and outer ($R>\Rp$) disk, respectively. We further define the \emph{trough} of the gap $R_\text{gap}^\text{min}$ as the radius within the gap where $\Delta\bar{\Sigma}/\Sigma_0$ is lowest, which is always located exterior to the planet's semimajor axis. Observing the highly asymmetric gap structure in Fig.~\ref{fig:lowmass-adb-gap-structure}, we finally define the \emph{trailing} gap edge as the narrow region just outside of the planet's coorbital zone where a steep negative radial surface density gradient forms due to gap opening. The inner edge of this region roughly coincides with the radial location where the planet's spiral waves steepen into shocks, or $R\approx\Rp+\xshock$ with $\xshock$ given by \citet{goodman-rafikov-2001}
\begin{equation}
    \label{eq:shock-width}
    \xshock \approx 0.93\left(\frac{\gamma+1}{12/5}\frac{\Mp}{\Mth}\right)^{-2/5} \Hp\Rightarrow \xshock\sim 0.03\,\Rp.
\end{equation}

This trailing gap edge is of key importance, as it is a region where a steep surface density gradient can form close to the planet due to gap opening. The vortensity contrast along this gap edge can then seed the Rossby wave instability, spawning vortices that refill the trough of the gap and therefore restore the outer, negative Lindblad torque as they dissipate. As the planet continues to carve a gap, the sharp vortensity gradient is reestablished, and the repeated formation and dissipation of vortices ultimately sustains inward migration \citepalias{MCNALLY-ETAL-2019A}.

In the next paragraph we describe this development for the adiabatic model. The radiative model is then discussed in Sect.~\ref{sub:radiative-runaway}.

\subsection{Turbulent migration and runaway episode in adiabatic run}
\label{sub:adiabatic-runaway}

\begin{figure*}
    \includegraphics[width=\textwidth]{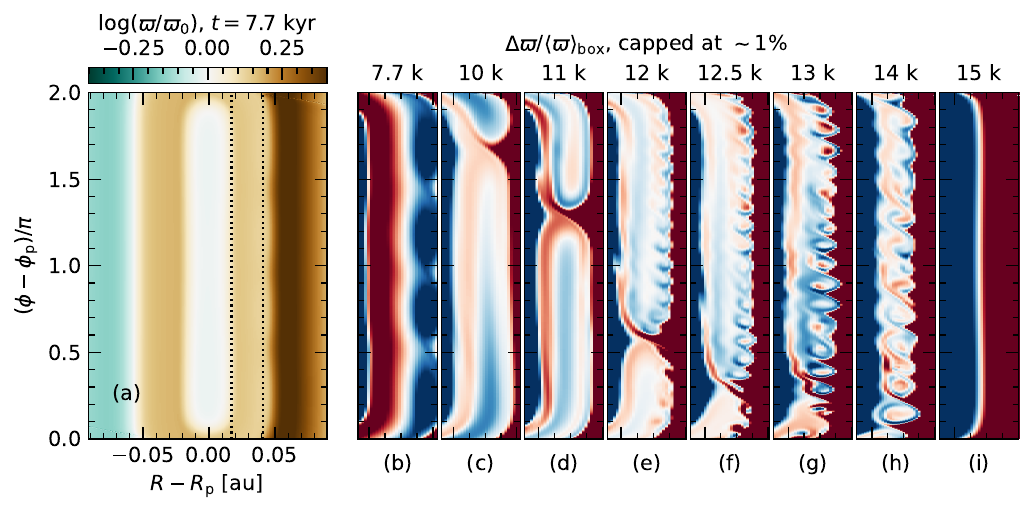}
    \caption{Panel \emph{a}: perturbed vortensity heatmap in the gap region for the adiabatic model at $t=7.7$\,kyr, right before the trailing gap edge ($R=\Rp+0.05$) becomes RWI unstable. Remaining panels: snapshots of the perturbed vortensity at the trailing gap edge with respect to the azimuthal median within the plotted region, intended to highlight the formation of vortices. The panels do not share the same color scale, but deviations are of order 1\%. Four vortices form initially (panel \emph{b}) and merge into one massive vortex (panels \emph{c}--\emph{d}), which breaks into $\approx20$ small-scale vortices (panels \emph{e}--\emph{h}) that remain active until the planet is pushed over the inner gap edge. The flow is smooth during the runaway episode (panel \emph{i}), after which the planet slows down and the process begins anew.}
    \label{fig:lowmass-adb-vortex-assisted}
\end{figure*}

As mentioned in the previous section, the steady carving of a gap by the planet leads to the formation of a sharp vortensity gradient at the trailing gap edge (see Fig.~\ref{fig:lowmass-adb-gap-structure}). This gap edge eventually becomes RWI-unstable, and vortex activity begins to appear at $t\sim7.7$\,kyr (see Fig.~\ref{fig:lowmass-adb-vortex-assisted}). Initially, four vortices form at the trailing gap edge, but subsequently a series of $\approx20$ small-scale vortices spawn at the same radial location. The intermittent formation, merging, and dissipation of these vortices can be observed in maps of the perturbed vortensity, which we show in panels \emph{b}--\emph{h} of Fig.~\ref{fig:lowmass-adb-vortex-assisted}. The vortices refill the trough of the gap, restoring the negative Lindblad torque and allowing the planet to continue its inward migration, until the planet eventually experiences a runaway episode at $t\sim16$\,kyr.

At the end of this relatively brief runaway episode, the planet arrives at a ``fresh'', unperturbed region of the disk, where the gap opening process begins anew: the planet establishes an asymmetric gap profile and temporarily slows down as a result, but the trailing gap edge eventually becomes RWI-unstable and vortex activity forces a sustained inward migration rate onto the planet. This behavior is identical to that observed in the model in Sect.~\ref{sec:comparison-to-mcnally}, and agrees with the findings of \citetalias{MCNALLY-ETAL-2019A} in the absence of radiative processes.

\subsection{Turbulent migration and runaway episode in radiative run}
\label{sub:radiative-runaway}

The radiative model shows a few similarities with the adiabatic run in that the overall migration pattern is qualitatively similar after an asymmetric gap has been established. However, this model also exhibits a few key differences that are worth discussing in more detail.

The first difference is the significantly longer phase of steady inward migration ($t\approx5$--25\,kyr) compared to the adiabatic model ($t\approx3$--14\,kyr). It is possible that this is related to the overall slightly faster migration in the radiative model, which would delay the onset of the RWI (and therefore a type-III episode) until the planet has slowed down enough for a sharp trailing gap edge to form. This faster migration could in turn be linked to the type-I-like behavior at $t\approx2$--5\,kyr and the time needed for the planet to smoothly transition to the feedback regime.

A second difference can be found at the time when the RWI becomes active. The latter can be seen as ``noise'' in the migration timescale in the bottom panel of Fig.~\ref{fig:migration-feedback} due to the presence of vortices introducing a stochastic component to the planet's migration track. Through this figure, we find that in the adiabatic model the vortex-assisted phase is clearly contained between $t\approx8$--14\,kyr. In contrast to this, the first signs of the RWI are weakly visible at $t\approx17$\,kyr in the radiative model, with several ``spikes'' of noise appearing until eventually the RWI becomes fully active between $t\approx22$--25\,kyr. This behavior is unlike what was reported by \citetalias{MCNALLY-ETAL-2019A}, and indicates the operation of a mechanism that suppresses the RWI in the radiative model.

\begin{figure*}
    \includegraphics[width=\textwidth]{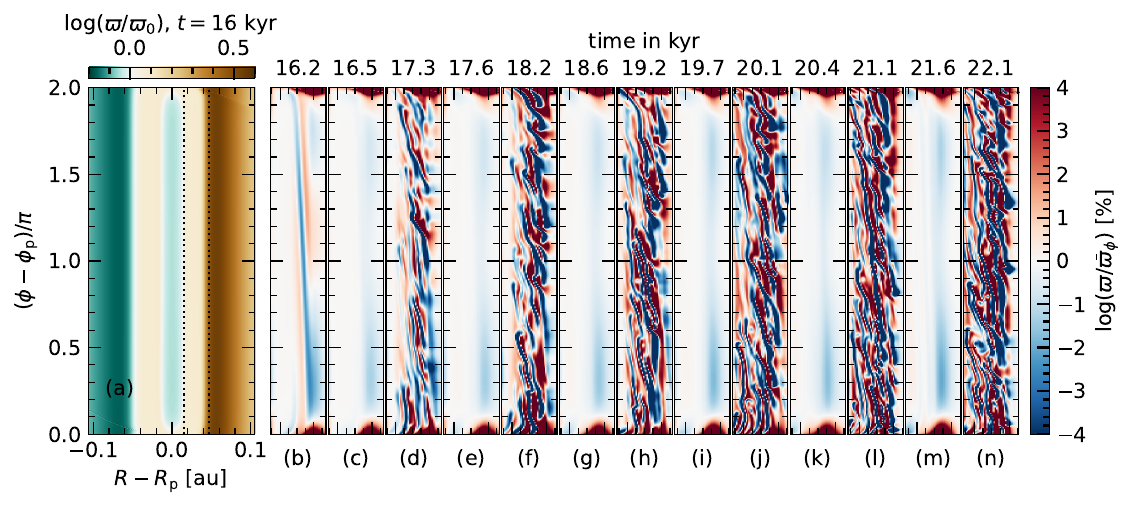}
    \caption{Snapshots of the perturbed vortensity at the trailing gap edge in the radiative model during the growth phase of the RWI. Panel \emph{a} shows the local disk structure, with the trailing gap edge enclosed by vertical dotted lines. Turbulence due to the RWI develops and subsides repeatedly between $t\in[16,25]$\,kyr, before the RWI becomes fully active at $t\approx25$\,kyr. By this time, the turbulent region extends to the planet's coorbital zone, influencing the planet's migration track.}
    \label{fig:lowmass-rad-vortex-suppression}
\end{figure*}

We first show that this is indeed the case by plotting several snapshots of the perturbed vortensity at the trailing gap edge in Fig.~\ref{fig:lowmass-rad-vortex-suppression} during $t\in[16,25]$\,kyr. This series of snapshots shows the repeated growth and decay of nonaxisymmetric features at the trailing gap edge between panels \emph{b}--\emph{m}, before the RWI activates permanently in panel \emph{n}. While the growth of such features is indicative of the RWI, their decay warrants further investigation.

\begin{figure}
    \includegraphics[width=\columnwidth]{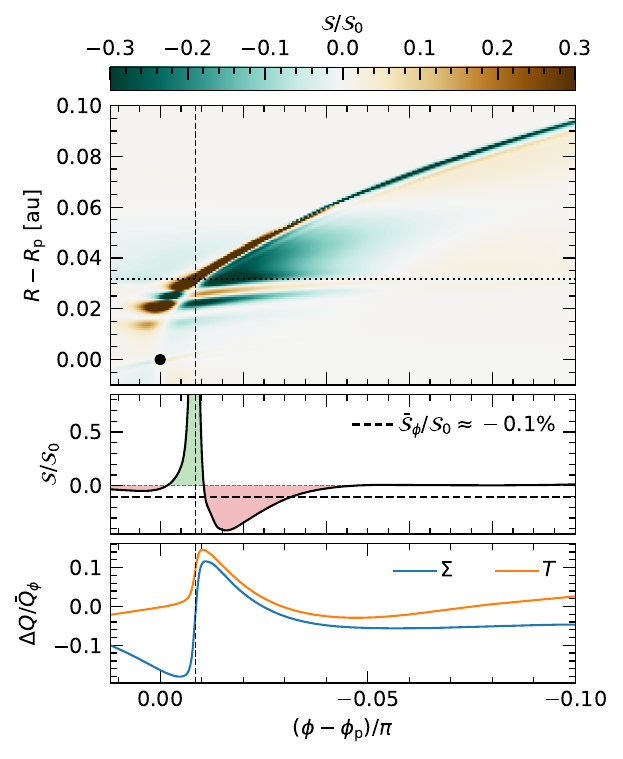}
    \caption{Top: map of the baroclinic forcing term $\mathcal{S}$ in the gap region at $t=18.2$\,kyr for the radiative model. 
    Middle: azimuthal slice of $\mathcal{S}$ at the location where the RWI is expected to activate (see Fig.~\ref{fig:lowmass-adb-vortex-assisted}), with a net negative forcing near the spiral shock. The horizontal dashed line is magnified by a factor of 100. Bottom: azimuthal surface density and temperature slices at the shock front. We normalize $\mathcal{S}$ using $\mathcal{S}_0 = \varpi_{0}^{R=1\,\text{au}}\,\OmegaK^{R=1\,\text{au}}$.}
    \label{fig:lowmass-rad-forcing}
\end{figure}

To understand this behavior, we compute the baroclinic forcing term $\mathcal{S}$ in the vortensity equation. By taking the curl of Eq.~\eqref{eq:navier-stokes-2}, we obtain
\begin{equation}
    \label{eq:vortensity-forcing}
    \D{\varpi}{t} = \mathcal{S} = \frac{1}{\Sigma^3}\,\nabla\Sigma\times\nabla P = \frac{P}{T\Sigma^3}\,\nabla\Sigma\times\nabla T.
\end{equation}
We then plot a map of $\mathcal{S}$ in the gap region at $t=18.2$\,kyr (during the RWI burst in panel \emph{f} of Fig.~\ref{fig:lowmass-rad-baroclinic}) in the top panel of Fig.~\ref{fig:lowmass-rad-forcing}, revealing a pair of peaks about the planet's spiral shock. The peaks are asymmetric, with a net negative baroclinic forcing (see middle panel of Fig.~\ref{fig:lowmass-rad-forcing}), resulting in a vortensity sink near the spiral shock.

By expanding the components of $\mathcal{S}$ in Eq.~\eqref{eq:vortensity-forcing}, we have
\begin{equation}
    \label{eq:vortensity-forcing-components}
    \mathcal{S} = \frac{PR^2}{\Sigma^2}\left(\DP{\log\Sigma}{\log R}\DP{\log T}{\phi} - \DP{\log \Sigma}{\phi}\DP{\log T}{\log R}\right).
\end{equation}
Since $\DP{\log T}{\phi}$ and $\DP{\log\Sigma}{\phi}$ are both positive with similar magnitude in the post-shock region (see bottom panel of Fig.~\ref{fig:lowmass-rad-forcing}), the sign of $\mathcal{S}$ will be decided by comparing $\DP{\log\Sigma}{\log R}$ to $\DP{\log T}{\log R}\approx q$ (since radiative cooling maintains the irradiation temperature profile of Eq.~\eqref{eq:initial-conditions}). At the trailing gap edge, the steep surface density gradient sets $\DP{\log\Sigma}{\log R} \ll q < 0$, such that overall $\mathcal{S} < 0$.

We note that, at the same time, the positive radial density gradient between $R_\text{gap}^\text{min}$ and $R_\text{gap}^\text{out}$ (see Fig.~\ref{fig:lowmass-adb-gap-structure}) results in $\DP{\log\Sigma}{\log R} > 0$ and $\mathcal{S} > 0$ there. Nevertheless, that region is neither RWI-unstable nor relevant to the problem of planet migration in this regime, as the planet has drifted inwards too far to drive gap opening into or feel a significant torque from that region.

We also highlight that this baroclinic forcing is inherently linked to treating radiative cooling in the disk. In our adiabatic model, where we force conservation of entropy across the shock front, Eq.~\eqref{eq:vortensity-forcing} automatically evaluates to zero. In addition, since the dominant term in Eq.~\eqref{eq:vortensity-forcing-components}, $\DP{\log\Sigma}{\log R}$, is coupled to azimuthal temperature variations, a locally isothermal model would not capture this effect either.

Of course, the vortensity sink discussed here directly competes with the vortensity sourcing due to gap opening, and the net effect between these two processes determines the overall behavior of the RWI. However, the vortensity sink term is driven by spiral shocks, which have more or less the same structure regardless of gap depth, while the gap itself can get arbitrarily deep as long as the planet is present. As a result, the RWI is suppressed only until a deep enough gap has been carved, at which point it becomes active and the planet eventually experiences a runaway episode.

A third difference between the adiabatic and radiative models is the faster migration observed in the latter immediately after the runaway episode at $t\approx 27$\,kyr. This is not surprising, however, as it is merely a repetition of the behavior observed during the first 2--5\,kyr of the simulation, where the formation of a shallow gap around the corotating region results in moderate vortex activity due to baroclinic forcing (see also Sect.~\ref{sub:feedback-regime} and Fig.~\ref{fig:lowmass-rad-baroclinic}). Nevertheless, this result reinforces the idea that migration in the feedback regime is cyclical in nature \citepalias[as shown by][]{MCNALLY-ETAL-2019A}.

\subsection{Section summary}
\label{sub:feedback-summary}

Overall, the presence of both a corotating region and a shallow gap in the feedback regime subjects planet migration to several different mechanisms that can drive baroclinic forcing when radiative cooling is considered. This can result in prolonged segments of smooth migration between runaway episodes, faster migration after such episodes, and complex vortex dynamics.

At the same time, however, the picture can be considered qualitatively similar between isothermal, adiabatic, and radiative models, in that the presence of a cycle between phases of feedback slowdown, vortex-assisted migration, and type-III episodes is maintained. As a result, while radiative cooling can delay the formation of vortices, it cannot prevent them from influencing the planet's migration track, and the planet ultimately experiences runaway episodes.

\section{The gap-opening regime: $\Mf<\Mth<\Mp$}
\label{sec:gap-opening-regime}

In the previous section we analyzed extensively the behavior of a super-Earth in the feedback regime, finding that the concept of the ``inertial limit'' originally suggested by \citet{hourigan-ward-1984}, \citet{ward-hourigan-1989}, and \citet{ward-1997a} is likely to not be applicable in the context of realistic disk conditions. This happens due to the turbulent diffusion induced by small-scale vortices at the planet's gap edge, consistent with the findings of \citetalias{MCNALLY-ETAL-2019A}. A natural next step is to then investigate the behavior of a super-Earth in the gap-opening regime, or $\Mp\gtrsim\Mth$, which the planet is expected to transition into as the thermal mass decreases inwards nearly inversely proportional to the distance from the star ($\Mth\propto R^{6/7}$ for $h\propto R^{2/7}$, see Eq.~\eqref{eq:thermal-mass}).

For our disk parameters, a planet with mass $\Mp=2\times10^{-5}\,\Msun=6.7\,\Mearth$ translates to $1.92\,\Mth$. We could then in principle repeat the above two models with this planet mass. However, we note that this planetary mass now lies quite close to the median mass of observed Kepler super-Earths, making it an appealing option to adapt our setup in hopes of probing the migration track of the ``typical'' super-Earth while still maintaining a connection to the previous sections. In light of the findings of the AGE-PRO ALMA large program\footnote[1]{Based on unpublished data, private communication.}, where the targeted class-II disks were found to have a median mass of $\lesssim 0.01\,\Msun$ and cutoff radius $R_\text{c}\sim 30$\,au, and assuming a surface density power-law $s=-1$ with an exponential taper at $R_\text{c}$, we find that the surface density at 1\,au around a solar-mass star should be $\Sigma_\text{ref}\approx 500\,\text{g}/\text{cm}^2$. We adopt this value for our models in this section. The feedback mass is then $\Mf\approx0.47\,\Mearth$, such that $\Mp\approx14\,\Mf$.

\begin{figure}
    \includegraphics[width=\columnwidth]{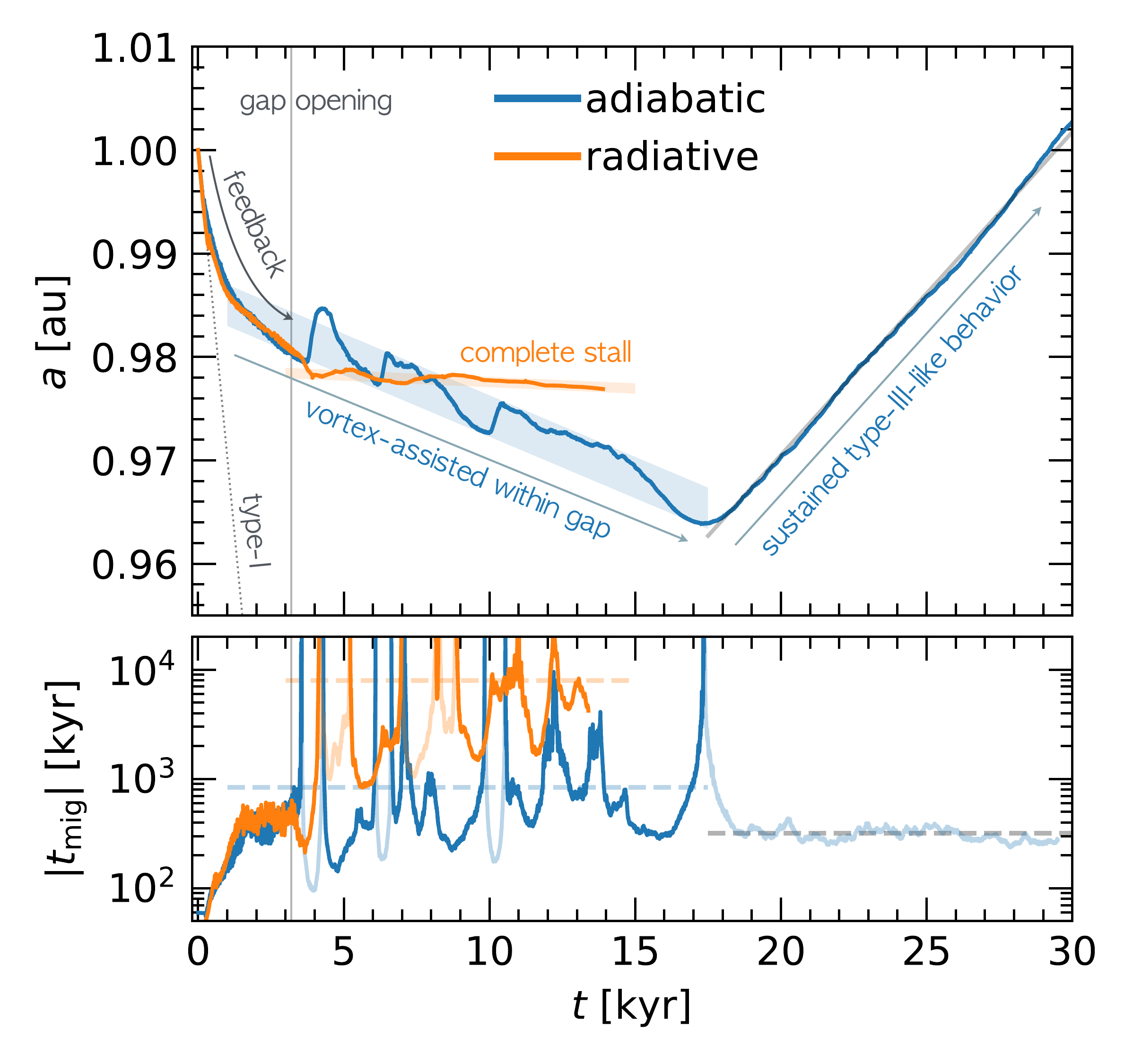}
    \caption{Migration tracks for a $6.7\,\Mearth$-mass planet in the gap-opening regime. After a brief period of type-I migration, the planet begins to carve a gap and slow down. The two models diverge after the gap profile has been established, with the planet trending inwards in the adiabatic model but completely halting in the radiative model. After $t\sim17$\,kyr the planet transitions to a steady outward migration phase in the adiabatic model.}
    \label{fig:migration-gap}
\end{figure}

Given that this $6.7\,\Mearth$-mass planet now exceeds the thermal mass, our immediate expectation is for significant gap opening to take place. Combined with the lower disk mass (i.e., a much smaller feedback mass), we anticipate that migration will be substantially slower due to both the weaker disk torques and gap opening. This is confirmed in Fig.~\ref{fig:migration-gap}, where we show the migration tracks for the adiabatic and radiative models, and which show a radial displacement of order $\sim0.02$\,au after 10\,kyr.

However, we also observe significant differences in the overall behavior between the two models, even though the two migration tracks appear functionally identical for the first 4\,kyr. The planet begins to carve a gap and transition into the feedback/gap-opening regimes as a result, and the two models diverge soon after the gap profile has been established. The planet continues to drift inwards until $t\sim17$\,kyr in the adiabatic model, before transitioning to a steady outward migration phase for the remainder of the simulation. In the radiative run, however, the planet completely halts beyond $t\approx4$\,kyr. This key difference in behavior is intimately linked to the gap opening process in disks with radiation hydrodynamics, and we will explore this in more detail later in Sect.~\ref{sub:gap-opening-stalling}. In the next subsections we analyze different phases of the migration process in both models.

\subsection{Inward drift phase in the adiabatic model}
\label{sub:inward-drift}

\begin{figure}
    \includegraphics[width=\columnwidth]{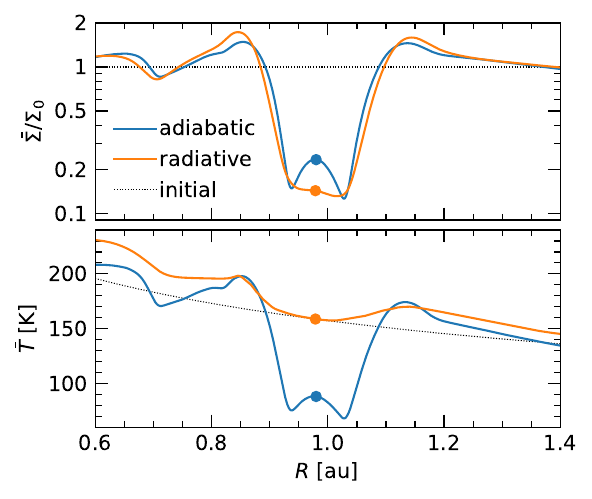}
    \caption{Gap structures at $t=3.8$\,kyr, right before the two models diverge. Top: azimuthally averaged surface density profiles for the adiabatic (blue) and radiative (orange) models, showing overall agreement but with more than double the amount of coorbital material in the adiabatic model. Bottom: radial temperature profiles for both models, showing a much colder gap region around the planet's location in the adiabatic model.}
    \label{fig:gaps-thermal}
\end{figure}

Within $\sim3.8$\,kyr a gap has been established around the planet's orbit, which we show in the top panel of Fig.~\ref{fig:gaps-thermal}. While the overall disk surface density structure looks quite similar between the two models, a noticeable difference is the presence of coorbital material in the adiabatic model as opposed to the flat surface density profile at the gap center in the radiative model, resulting in an effectively shallower gap in the former. This "double trough" gap structure is expected for planets embedded in isothermal or adiabatic disks \citep{miranda-rafikov-2020a} and is a consequence of the planet-induced spiral shocks depositing angular momentum approximately a shock distance $\xshock$ away (see Eq.~\eqref{eq:shock-width}). We refer the reader to \citet{cordwell-rafikov-2024} for a detailed analysis of the early stages of gap opening in isothermal disks, where very similar features are observed.

Importantly, this "double trough" structure implies that a substantial amount of coorbital material is trapped in the corotating region, such that planet--vortex interactions can efficiently refill the gap and assist inward migration. This behavior is similar to the mechanism described in \citetalias{MCNALLY-ETAL-2019A} for planets in the feedback regime, but is also reminiscent of the "vortex-driven" migration discussed for massive planets in \citet{lega-etal-2021,lega-etal-2022}, with the key difference being the presence of several small vortices as opposed to a single, massive vortex at the gap edge.

The bottom panel of Fig.~\ref{fig:gaps-thermal} shows the radial temperature profile for both models, where it is evident that the two differ significantly. In the radiative model, the temperature is overall moderately higher compared to the initial profile due to the dissipation of planet-driven shocks contributing to disk heating \citep{rafikov-2016,ziampras-etal-2020a}. In the adiabatic model, however, where neither shock heating nor a background radiative equilibrium state are captured, the temperature profile simply adjusts due to adiabatic compression and expansion around gap opening regions. This results in a significantly colder gap region around the planet's location and near the secondary gap forming at $R\approx0.7$\,au, and a moderately warmer disk elsewhere. We therefore stress that the adiabatic model does not yield a good representation of the disk's thermal structure even though the disk was initially optically thick, highlighting the importance of including radiative processes in this context.

\begin{figure*}
    \includegraphics[width=\textwidth]{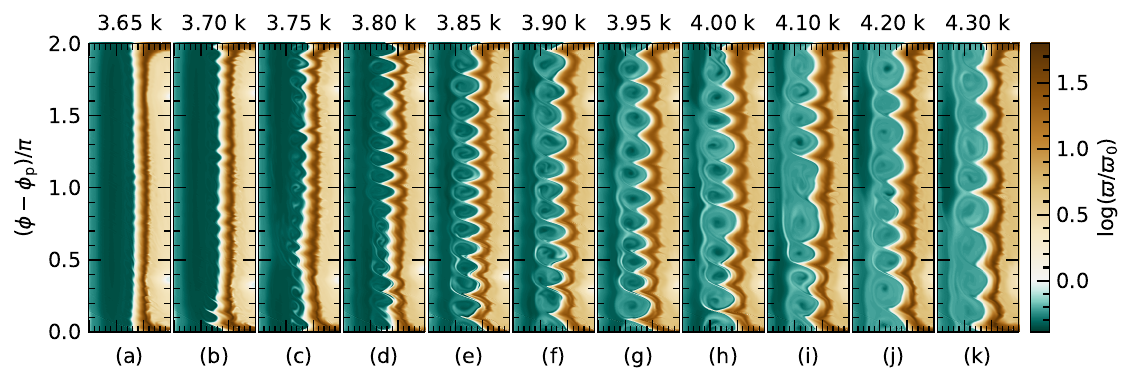}
    \caption{A series of snapshots of the perturbed vortensity at the interface between the inner gap edge and the coorbital region in the adiabatic model around the time of the first vortex burst at $t\sim$4\,kyr. The vortices grow (panels \emph{a}--\emph{d}) and merge into larger vortices (panels \emph{e}--\emph{h}) before they start decaying (panels \emph{i}--\emph{k}).}
    \label{fig:thermal-adb-vortex-first}
\end{figure*}

Planet--vortex interactions are observed in our simulations as soon as the planet approaches its inner gap edge and a sharp vortensity contrast between the gap edge and the corotating region is established. This results in a burst of vortex activity, with $\sim$20 vortices growing along the interface between gap edge and corotating region. The vortices are short-lived, but the turbulent diffusion they induce results in partial refilling of the gap, reinvigorating the positive inner Lindblad torque and pushing the planet outwards (see also Fig.~\ref{fig:migration-gap}). The exact same behavior is then observed as the planet subsequently approaches the outer gap edge, with a similar burst of vortices replenishing the negative outer Lindblad torque and reversing the planet's migration direction once again. Over the course of $\sim13$\,kyr, this cycle repeats several times (see shaded blue region in Fig.~\ref{fig:migration-gap}), with the planet ``bouncing'' between the inner and outer gap edges, each bounce refilling the gap and partially restoring the Lindblad torques. Given that the net Lindblad torque is negative, this process results in the planet trending inwards on average, with a migration timescale of $\sim 0.8$\,Myr (see lower panel of Fig.~\ref{fig:migration-gap}).

\begin{figure*}
    \includegraphics[width=\textwidth]{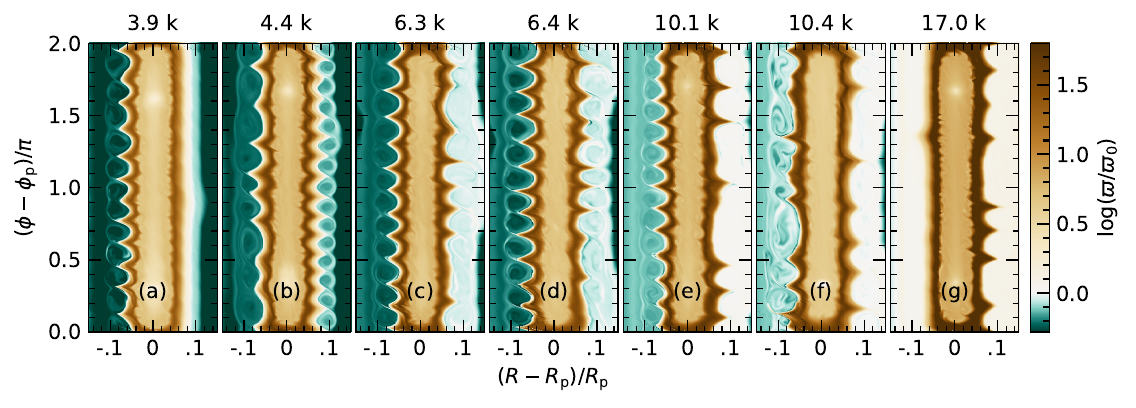}
    \caption{Similar to Fig.~\ref{fig:thermal-adb-vortex-first}, but showing the perturbed vortensity across the entire gap region and over the course of the inward migration phase highlighted in blue on Fig.~\ref{fig:migration-gap}. The planet interacts with its inner gap edge in panels \emph{a}, \emph{c}, and \emph{e}, and with its outer gap edge in panels \emph{b}, \emph{d}, and \emph{f}, eliminating the vortensity contrast between the gap edges and the corotating region in the process. This contrast is completely erased by $t\sim17$\,kyr (panel \emph{g}).}
    \label{fig:thermal-adb-vortex-repeated}
\end{figure*}

In Fig.~\ref{fig:thermal-adb-vortex-first} we show snapshots of the perturbed vortensity heatmaps for the adiabatic model around the time of the first vortex burst at $t\sim$3.8\,kyr, focusing on the interface between the gap edge (in blue) and the corotating region (in brown). The series of panels showcases the formation of $\sim$20 vortices that grow and merge into $\sim$12 larger vortices before decaying over $\sim1000$~orbits. Their dissipation mixes material between the gap edge and corotating region, a process that is visible by comparing the color contrast across the interface of the two regions between panels \emph{a} and \emph{k}.

In Fig.~\ref{fig:thermal-adb-vortex-repeated} we then show the same quantity across the entire gap region and over the course of the inward migration phase highlighted in blue on Fig.~\ref{fig:migration-gap}. This set of snapshots illustrates the repeated vortex activity at the gap edges, each panel corresponding to a timestamp where the planet reverses its direction. The planet switches from inward to outward migration in panels \emph{a}, \emph{c} and \emph{e} due to vortex bursts at the inner gap edge, and from outward to inward migration in panels \emph{b}, \emph{d} and \emph{f} due to bursts at the outer gap edge. Since each burst mixes material between the gap edge and corotating region (see also Fig.~\ref{fig:thermal-adb-vortex-first}), this repeated cycle of vortex bursting and decay effectively eliminates the low-vortensity bands at the gap edges, rendering them quasi-stable against the RWI and preventing future vortex bursts from occurring.

It is important to note that a vortex burst results in rapid migration in the opposite direction, triggering a second burst at the opposite gap edge. This results in pairs of vortex bursts separated by 100--400~years. Given that the first burst happens at the inner gap edge, this behavior is the reason why vortices are still visible on the inner gap edge during a vortex burst at the outer gap edge.

\subsection{Gap opening and stalling in the radiative model}
\label{sub:gap-opening-stalling}

\begin{figure*}
    \includegraphics[width=\textwidth]{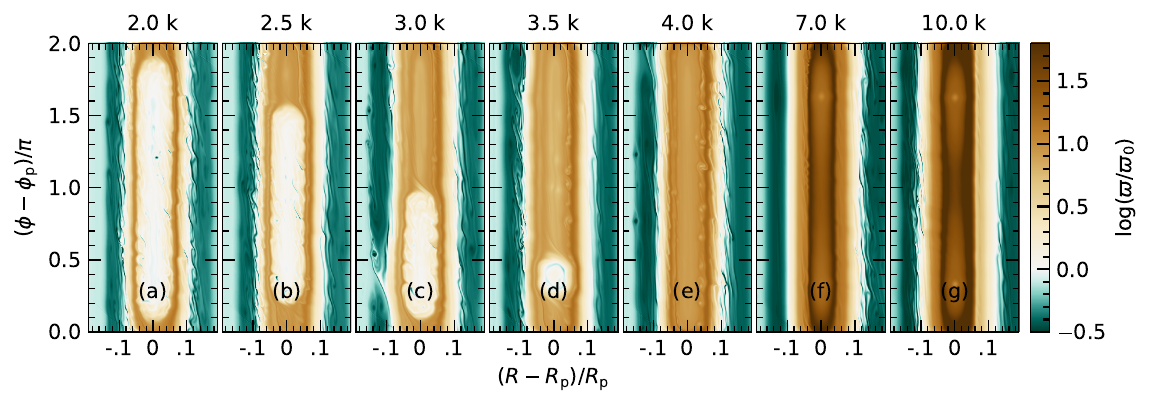}
    \caption{Snapshots of the perturbed vortensity similar to Fig.~\ref{fig:thermal-adb-vortex-repeated} for the radiative model. Using the vortensity as a proxy for the inverse of the surface density (i.e., darker orange colors denote a deeper gap), these panels follow the gap opening process and the eventual double trough structure at $t\sim7$\,kyr. At the same time, the maps reveal a multitude of small-scale, corotating vortices and turbulent gap edges.}
    \label{fig:thermal-rad-gap}
\end{figure*}

While an inward drift phase is observed in the adiabatic model, the radiative model shows a different behavior once a gap profile has been established by $t\sim4$\,kyr. Contrary to the vortex-rich environment discussed in the previous section, the planet in the radiative model clears a deep gap, empties its coorbital region, and practically grinds to a halt, as seen in Fig.~\ref{fig:migration-gap}. Indeed, in the series of snapshots in Fig.~\ref{fig:thermal-rad-gap}, high-vortensity bands at both gap edges are visible, but as the planet does not get to approach either gap edge closely, no vortex bursts are observed.

At the same time, using the vortensity as a proxy for the inverse surface density, the same series of panels follows the process of a deep gap being carved around the planet's orbit, with the coorbital region being emptied of material within 2\,kyr, in stark contrast to the adiabatic model. Given that the planet is migrating inwards, horseshoe orbits form a tadpole-shaped region ahead of the planet, resulting in that region being cleared last. Interestingly, as gap opening continues, a double trough structure is observed once again at $t\sim7$\,kyr in the vortensity maps in Fig.~\ref{fig:thermal-rad-gap}.

\begin{figure}
    \includegraphics[width=\columnwidth]{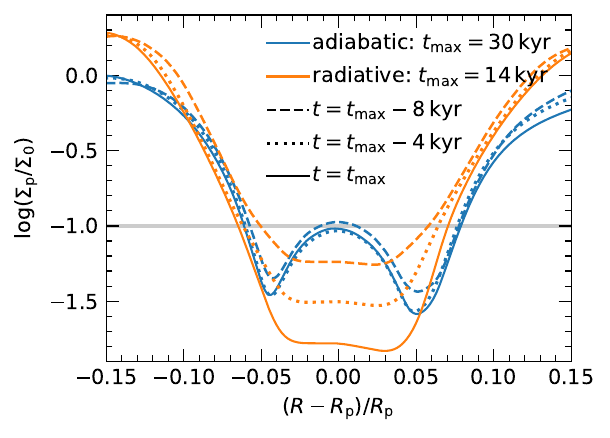}
    \caption{Azimuthally averaged perturbed surface density centered on the planet's location for three different timestamps in the adiabatic and radiative models. The gap profile remains stagnant at $\Sigma(\Rp)/\Sigma_0\approx10\%$ for the adiabatic model, but continues to deepen past that in the radiative model.}
    \label{fig:gap-progression}
\end{figure}

\begin{figure}
    \includegraphics[width=\columnwidth]{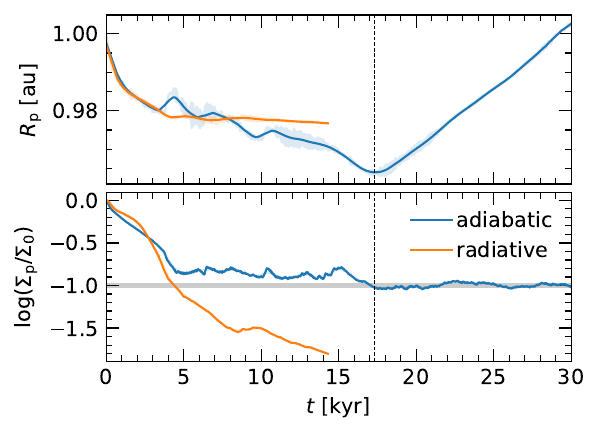}
    \caption{Time evolution of the ``gap depth'' $\Sigma(\Rp)/\Sigma_0$, showing that the gap profile remains stagnant for extended periods of time in the adiabatic model but continues to deepen in the radiative run. The migration tracks are shown for reference.}
    \label{fig:gap-depth-time}
\end{figure}

To quantify the differences in the gap structure between the two models, we plot the azimuthally averaged surface density with respect to its initial profile around the planet's radial location $\Rp$ at three different timestamps in Fig.~\ref{fig:gap-progression}. This figure shows that the gap profile remains stagnant at $\Sigma(\Rp)/\Sigma_0\approx10\%$ for the adiabatic run over the last $\sim8$\,kyr of the simulation, but continues to deepen well below that value in the radiative model. A double-trough structure in the adiabatic model is also clearly visible in this figure. Using $\Sigma(\Rp)/\Sigma_0$ as a metric for the gap depth, we also track this quantity as a function of time for both models and plot it side-by-side with the two migration tracks in Fig.~\ref{fig:gap-depth-time}. Here, it becomes clear that the gap depth stagnates during both the vortex-rich inward drift phase and the steady outward migration phase in the adiabatic model, but continues to deepen in the radiative model.

The behavior described above is fully consistent with the findings of \citet{miranda-rafikov-2020a}, \citet{ziampras-etal-2020b,ziampras-etal-2023a}, and \citet{zhang-zhu-2020}, who investigated planet-driven gap opening in disks with cooling. As the cooling timescale approaches unity in the gap region, the double trough structure expected for adiabatic or isothermal models gives way to a single, deep gap. This happens due to radiative damping of spiral shocks, which results in angular momentum deposition much closer to the planet's orbit than the shocking distance $\xshock$ in Eq.~\eqref{eq:shock-width}. In the context of our simulations, even though the disk is initially optically thick and adiabatic, the partial gap opening induced by the planet reduces the cooling timescale within the gap region, focusing the angular momentum deposition closer to the planet's orbit and resulting in a deeper gap, which further reduces the cooling timescale. This feedback loop accelerates the gap opening process, clears the coorbital region of material, and halts the planet's migration.

\begin{figure}
    \includegraphics[width=\columnwidth]{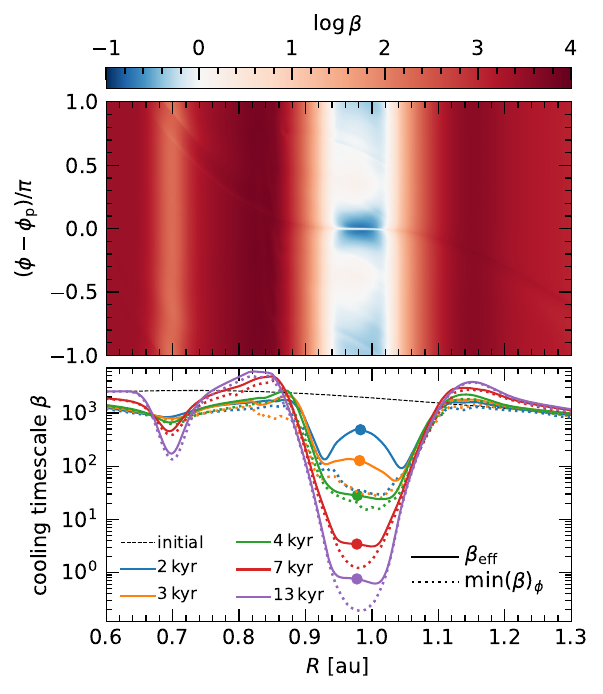}
    \caption{Top: the cooling timescale $\beta$ through Eq.~\eqref{eq:cooling-timescale} at $t=13$\,kyr for the radiative model. Bottom: radial profiles of $\beta$ using different metrics, describing cooling near spiral shocks ($\min(\beta)_\phi$) and the gap as a whole ($\beta_\text{eff}$). By $t\sim 10$\,kyr, the cooling timescale is of order unity or below within the gap.}
    \label{fig:beta-radiative}
\end{figure}

In the top panel of Fig.~\ref{fig:beta-radiative} we show a heatmap of the cooling timescale $\beta$ at $t=13$\,kyr, with $\beta$ computed following \citet{ziampras-etal-2023b} as
\begin{equation}
    \label{eq:cooling-timescale}
    \beta = \frac{1}{1+f}\frac{e}{|\Qcool|}\OmegaK,\qquad f = 16\pi\frac{\tau\,\taueff}{6\tau^2+\pi},
\end{equation}
which appropriately combines the effects of surface and in-plane cooling in Eqs.~\eqref{eq:source-terms-3}~\&~\eqref{eq:source-terms-4}. From this panel it becomes clear that the cooling timescale is of order unity within the gap region, and even drops to a minimum of $\approx0.2$ in the planet's vicinity, where spiral shocks are excited.

To showcase the evolution of $\beta$ over time, we show in the lower panel of Fig.~\ref{fig:beta-radiative} two metrics for the cooling timescale in the gap region: the azimuthal minimum $\min(\beta)_\phi$, which roughly corresponds to $\beta$ in the vicinity of the planet and which should relate to the gap opening efficiency, and an ``effective'' cooling timescale for each annulus defined as
\begin{equation}
    \label{eq:effective-cooling-timescale}
    \beta_\text{eff}(R) = \frac{\oint_\phi e(R,\phi)\,\mathrm{d}\phi}{\oint_\phi \frac{e(R,\phi)}{\beta(R,\phi)}\,\mathrm{d}\phi},
\end{equation}
which more accurately reflects how efficiently the gap cools as a whole. Both metrics show that the cooling timescale drops from the initial value of $\sim10^3$ to $\sim1$ over the course of the simulation, reaching a minimum of $\sim0.2$ at $t\sim13$\,kyr. This also explains the eventual double trough structure reemerging at $t\sim7$\,kyr (see Fig.~\ref{fig:thermal-rad-gap}), as $\beta$ drops below unity and cooling inside the gap tends towards the isothermal limit.

\subsection{Outward migration in the adiabatic model}
\label{sub:outward-migration}

\begin{figure}
    \includegraphics[width=\columnwidth]{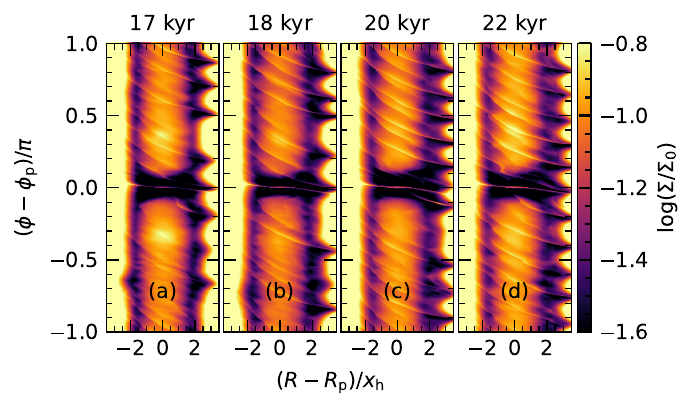}
    \caption{Heatmaps of the surface density around the planet's orbit in the adiabatic model during and after the migration reversal at $t\sim17$\,kyr. The pair of vortices in the corotating region merge and develop an asymmetry that results in a net positive torque on the planet, reversing the migration direction.}
    \label{fig:thermal-adb-outward}
\end{figure}

After several interactions between the planet and its gap edges in the adiabatic model, the high-vortensity bands on the gap edges are erased and the process of vortex bursts subsides. The planet then transitions into a steady outward migration at $t\sim17$\,kyr, as seen in Fig.~\ref{fig:migration-gap}. We find that this behavior is similar to that observed by \citetalias{MCNALLY-ETAL-2019A} for planets in the thermal mass regime (see curve `r8' in their Fig.~3): a pair of vortices form within the corotating region, and their merging develops an asymmetry that results in a net positive torque on the planet, reversing the migration direction. To confirm this, we show heatmaps of the surface density around the planet's orbit in Fig.~\ref{fig:thermal-adb-outward} during and after the migration reversal at $t\sim17$\,kyr.

Over the course of this outward migration phase, however, this asymmetry diminishes, yet the planet continues to migrate outwards at a steady rate. We speculate that given the very slow type-II migration rate expected for our value of $\alpha$ ($t_\text{visc}\approx 215$\,Myr as opposed to the observed $t_\text{mig}\approx 300$\,kyr), the planet is likely experiencing a type-III-like migration phase, where a net flow through the gap region is established and exerts a net dynamical corotation torque on the planet \citep{masset-papaloizou-2003,peplinski-etal-2008}. This argument is consistent with the observation that the surface density profile around the planet's orbit remains stagnant during this phase rather than the gap deepening over time, as shown in Figs.~\ref{fig:gap-progression}~\&~\ref{fig:gap-depth-time}.

Nevertheless, we note that this entire phase of outward migration is likely an artifact of our disk model, as the adiabatic approximation does not capture the gap opening process accurately. Perhaps ironically, the isothermal models of \citetalias{MCNALLY-ETAL-2019A} are both more self-consistent and at least applicable in the rapidly-cooling regions of the outer disk, even though the isothermal approach is more simplistic compared to the adiabatic model.

\subsection{Section summary}
\label{sub:gap-opening-summary}

Our analysis in the gap opening regime showed that radiative cooling is critical for modeling planet migration, with our radiative model behaving distinctly differently from the adiabatic run. In the radiative model, unlike in isothermal or adiabatic models, migration stalls. This happens due to the fact that planet-driven gap opening is strongly dependent on the local cooling timescale, such that the resulting gap structure and therefore torque balance are significantly altered.

Our adiabatic results, while no longer applicable once a deep gap has formed, are consistent with the findings of \citetalias{MCNALLY-ETAL-2019A}, where the presence of small-scale vortices at the gap edges can drive planet migration. In this type-II-like, vortex-assisted regime, the planet migrates much faster than the classic viscous type-II rate due to the presence of vortices but also much slower than the type-I rate due to the carving of a gap.

\section{A map of migration regimes}
\label{sec:map}

The focal point of our work has been to explore the different migration regimes that a super-Earth-mass planet can experience in a nearly inviscid, radiative disk. While in this work we address the feedback and gap-opening regimes, our results can be combined with previous studies to create a map of the different migration regimes in such disks, similar to the one presented by \citetalias{MCNALLY-ETAL-2019A} for isothermal disks.


\begin{figure*}
    \includegraphics[width=\textwidth]{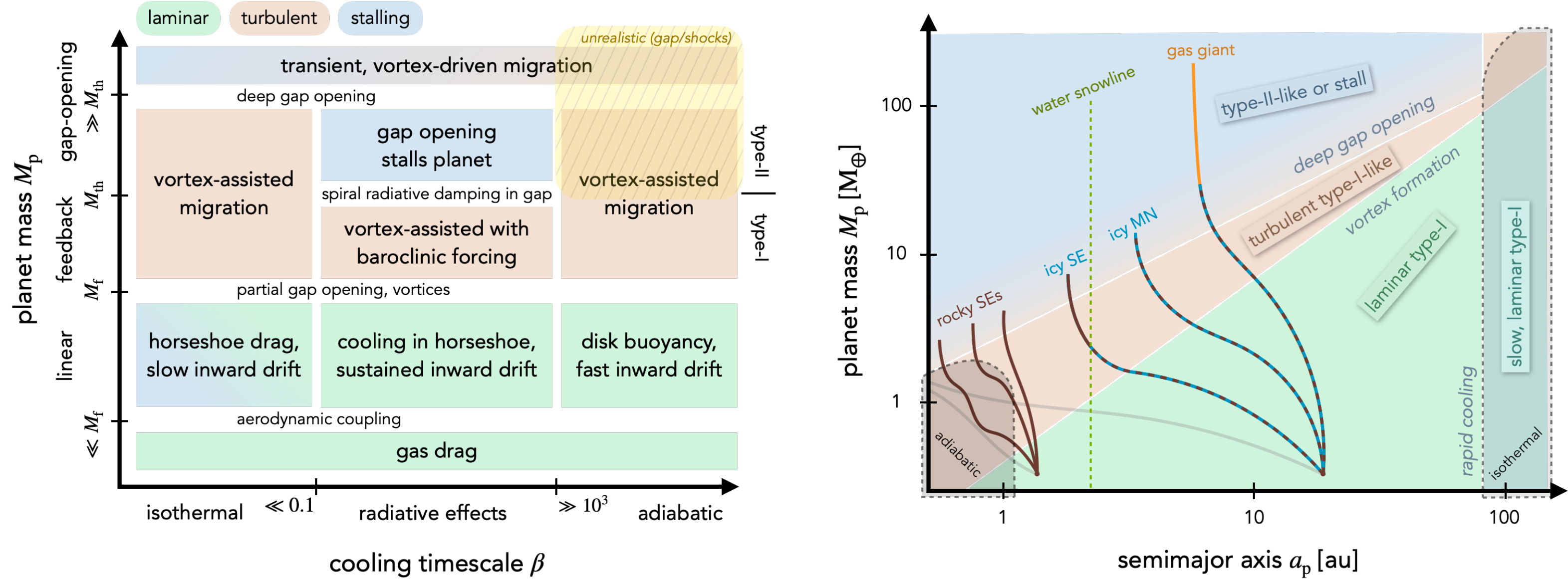}
    \caption{Left: conceptual ``city map'' of the different migration regimes for nearly inviscid disks $\alpha\lesssim10^{-5}$ as a function of the planet's mass $\Mp$ and the dimensionless cooling timescale $\beta$. The left side mirrors the work of \citetalias{MCNALLY-ETAL-2019A} in isothermal disks, and the results discussed in this work straddle the transition between type-I and type-II regimes. The map only offers a qualitative overview without accounting for MHD, dust--gas interaction, or other processes, and the exact boundaries between different regimes are not necessarily sharp. Right: sketch of possible formation tracks for planets interior or exterior to the water snowline based on the regimes identified in the left panel. Depending on the planet's initial location and its accretion efficiency, it is possible to form both rocky and icy super-Earths (`SE'), mini-Neptunes (`MN'), and gas giants. Their migration then stalls due to gap opening. The implications of radiative cooling in gap opening, while shown as a rather small blue box for $\Mp\gtrsim\Mth$ in the left panel, actually determine every planet's stalling location in the right panel.}
    \label{fig:butcher-map}
\end{figure*}

In the left panel of Fig.~\ref{fig:butcher-map} we present a ``city map'' of the different migration regimes for nearly inviscid disks ($\alpha\lesssim 10^{-5}$) as a function of the planet's mass $\Mp$ and the dimensionless cooling timescale $\beta$. The map is divided along the $x$ axis into three regimes, with the isothermal regime (left, $\beta\ll0.1$) mirroring the panels in the map of \citetalias{MCNALLY-ETAL-2019A}, the adiabatic regime (right, $\beta\gg10^3$) representing the optically thick, radiatively inefficient disk regions, and the radiative regime (center) covering the bulk of the disk where radiative processes can be important to planet--disk interaction. Along the $y$ axis, the map is further divided into five regimes based on the planet's mass relative to different thresholds that we have identified in this work. In the following paragraphs, we outline the different regimes in the map.

\subsection{Gas drag regime}
\label{sub:map-gas-drag}

Sufficiently low-mass objects which orbit at Keplerian speeds interact with the gaseous disk purely through gas drag, as the pressure-supported gas (typically) exerts a headwind onto them and causes them to drift inwards \citep{weidenschilling-1977}. This radial drift depends on the Stokes number of such objects (see Eq.~\eqref{eq:stokes}) with the radial velocity given by \citep[e.g.,][]{takeuchi-lin-2002}
\begin{equation}
    \label{eq:radial-velocity}
    u_{R,\text{d}} \approx \frac{1}{\St+\St^{-1}}\D{\log P}{\log R}\uK,
\end{equation}
and operates on $\lesssim$\,kyr timescales for mm-sized bodies at 1\,au, efficiently delivering them to the central star in the absence of pressure traps \citep[e.g.,][]{pinilla-etal-2012a,pinilla-etal-2012b}. As the size and therefore Stokes number of the object increases, the latter detaches from the gas and can be considered a ``planet'' for our dynamical purposes. Since $u_{R,\text{d}}$ is proportional to the pressure gradient, it does not inherently depend on the cooling timescale unless secondary processes such as snowline-related effects \citep[e.g.,][]{owen-2020} or self-shadowing instabilities \citep[e.g.,][]{dullemond-2000,wu-lithwick-2021} are considered, but such processes are not immediately relevant in our context.

\subsection{Linear regime}
\label{sub:map-linear}

Low-mass planets are expected to migrate in the type-I regime, which in the context of nearly-inviscid disks is characterized by a balance between the negative Lindblad torque \citep{goldreich-tremaine-1980} and the (typically positive) dynamical corotation torque \citep[DCT,][]{paardekooper-2014}. The latter is expected to scale with the vortensity contrast between the corotating region ($\varpi_\text{h}$) and the background disk ($\varpi_\text{p}$) \citep{mcnally-etal-2017}
\begin{equation}
\label{eq:dct}
\Gamma_\mathrm{DCT} = 2\pi \left(1-\frac{\varpi_\text{p}}{\varpi_\mathrm{h}}\right) \Sigma_\mathrm{p} \Rp^2 \xh \Omega_\mathrm{p}\left(\D{\Rp}{t} - u_{R,\text{g}}\right),
\end{equation}
which typically increases as the planet continues to migrate inwards on the condition that $\varpi_\text{h}$ is conserved. This is achieved in isothermal disks and in the absence of diffusion, and results in the planet drastically slowing down as it migrates inwards \citep{paardekooper-2014}, such that it would instead migrate at $\sim$\,Myr timescales. However, several studies have shown that the DCT can be significantly modified when baroclinic processes are considered.

In radiative disks, \citet{ziampras-etal-2024a} showed that vortensity can be efficiently generated in the corotating region for cooling timescales of the same order as the U-turn timescale of horseshoe orbits near the planet, such that the DCT is noticeably weakened and type-I-like migration can be sustained \citep[see also][for similar results in viscous disks]{pierens-2015}. This process can operate for $\beta\sim$1--100, covering a wide radial extent of the disk and efficiently delivering low-mass planets from the outer disk to the few-au regions.

For radiatively inefficient (i.e., adiabatic) disks, corresponding to the sub-au region, the disk buoyancy response to an embedded planet generates an array of linear waves that can both exert a significant torque on the planet \citep{zhu-etal-2012} but also drive vortensity generation within the planet's horseshoe region as they dissipate \citep{mcnally-etal-2020}. Admittedly, this process requires a cooling timescale much longer than the local disk buoyancy frequency, such that it quickly shuts down at the few-au scale \citep{yun-etal-2022}. Nevertheless, for a passively irradiated disk model at 1--2\,au, \citet{ziampras-etal-2024b} have shown that buoyancy-related torques are efficient enough to even reverse the DCT, propelling planets inwards.

Overall, in our context of nearly inviscid disks, type-I migration typically points towards the central star. While it could be argued that the outer regions of the disk can cool efficiently enough to represent the isothermal regime, we note that studies have shown that the low gas and dust densities in these region can instead render disk cooling inefficient \citep[e.g.,][]{bae-etal-2021}. Furthermore, even if an efficiently-cooling region existed somewhere in the disk, it would likely be unstable to the vertical shear instability \citep{nelson-etal-2013} which could sustain a turbulent $\alpha\gtrsim10^{-5}$ even in the presence of a planet \citep{stoll-etal-2017b,ziampras-etal-2023vsi}, in turn limiting the conservation of $\varpi$ in the corotating region.

\subsection{Feedback regime}
\label{sub:map-feedback}

Once the planet's mass exceeds the feedback mass in Eq.~\eqref{eq:thermal-mass}, the modification of the local disk environment by the planet can feed back onto its migration. In this regime, \citetalias{MCNALLY-ETAL-2019A} have shown with an extensive suite of isothermal models that the planet consistently forms a trailing gap edge that is RWI-unstable, spawning numerous small-scale vortices that sustain the planet's inward migration by diffusively refilling the gap and therefore maintaining a negative Lindblad torque from the outer disk. While this behavior is sensitive to the numerical approach, they showed that an increase in grid resolution helps resolve more small-scale vortices and therefore increases the migration rate. The result is a type-I-like, ``vortex-assisted'' migration regime that shows no signs of stopping and can operate for planets with masses up to a few times the feedback mass. We note that occasional episodes of type-III runaway migration are often observed in this feedback regime.

Our findings in Sects.~\ref{sec:comparison-to-mcnally}~\&~\ref{sec:feedback-regime} are fully consistent with this picture, especially in the adiabatic regime which behaves identically to the isothermal case. With the addition of radiative processes, however, the disk is subject to baroclinic forcing that slightly modifies this picture. Initially, numerous small-scale vortices are generated during the formation and saturation of the horseshoe region, in a process similar to that described by \citet{ziampras-etal-2024a}. This allows the planet to maintain a faster, type-I migration rate for slightly longer before transitioning to the feedback regime. There, radiative cooling along the planet-driven shock fronts acts as a vortensity sink, delaying the formation of vortices and therefore the onset of the vortex-assisted regime. Nevertheless, the RWI is eventually triggered and the planet migrates inwards, albeit with less frequent type-III episodes. 

Taking into account the above information, we expect that planet migration transitions from a laminar to a vortex-assisted, turbulent type-I-like regime for planets above the feedback mass, with qualitatively similar behavior across different thermodynamical models in that an inward drift is maintained. While this implies that radiative cooling does not fundamentally change the behavior of migrating planets in this regime, it does carry implications for the interpretation of planet-driven substructures in the outer disk \citep[][Meiners et~al., in prep.]{MCNALLY-ETAL-2019A}.

\subsection{Gap-opening regime}
\label{sub:map-gap-opening}

Given that in both the linear and feedback regime the planet continues to migrate inwards, identifying a means to stall the planet becomes crucial to prevent it from reaching the central star. In traditionally viscous disks this can be achieved by the planet carving a deep gap in the disk and transitioning to the type-II regime, slowing down its migration to match the viscous evolution of the disk \citep{lin-papaloizou-1986,ward-1997b}. In the absence of turbulence, this should translate to the planet stalling, as the disk no longer evolves viscously. The lack of viscous diffusion also means that even low-mass planets can open a deep gap, as long as they exceed the thermal mass (see Eq.~\eqref{eq:thermal-mass}).

However, gap opening cannot continue indefinitely. The lack of viscosity means that the steep gap edges shaped by a super-thermal-mass planet are easily RWI-unstable, leading to the formation of vortices that continue to drive the planet inwards. This process is conceptually similar to that detailed in the feedback regime, but with the added complication of the planet's gap opening process. In a way, planets in this regime migrate in a type-II analog of the feedback regime, in that the planet lies in a deep gap, but with a migration rate much higher than that expected by viscous evolution, if any takes place. Both our results and those of \citetalias{MCNALLY-ETAL-2019A} support this picture of a ``vortex-assisted'', type-II-like migration regime for super-thermal-mass planets in both the isothermal and adiabatic regimes, with typical migration timescales of the order of $\sim$\,Myr. We therefore make the distinction between the type-I and type-II regimes in our map based on the planet's mass relative to the thermal mass, but note the decisive effect of vortices in both regimes.

This behavior changes dramatically when radiative effects are included in the model. The interplay between the gap opening process and radiative cooling generates a feedback loop where the clearing of a partial gap reduces the cooling timescale in the gap opening region, ``focusing'' spiral angular momentum deposition closer to the planet and resulting in more efficient gap opening. This process operates on timescales much shorter than the planet's migration timescale, and as a result a deep gap is formed with the planet stalling at its center. The underlying mechanism has been studied in detail by \citet{miranda-rafikov-2020a,miranda-rafikov-2020b}, \citet{zhang-zhu-2020}, and \citet{ziampras-etal-2020b,ziampras-etal-2023a} for planets on fixed orbits, but naturally extends to migrating planets as long as they are massive enough to carve a gap on a timescale shorter than their migration timescale (i.e., $\Mp\gg\Mf$). Ultimately, the result is a new regime of type-II-like migration where vortices cannot operate close to the planet due to the deep gap, and the planet stalls rather than migrating in a vortex-assisted fashion.

It is worth highlighting that the regime described here is relevant as long as the cooling timescale in the gap region can drop below $\sim100$, triggering the feedback loop. This makes this regime applicable to a wide radial range of the disk, as the cooling timescale is expected to drop with radius and our models suggest that the mechanism is already active at 1\,au for a typical disk model. A more stringent requirement instead is that the planet exceeds the thermal mass, which for our model is $\{3.5,9,25,65\}\,\Mearth$ at $\{1,3,10,30\}$\,au and therefore would allow $\sim5$--$10\,\Mearth$ mass planets to stall at the $\sim$au scale.

\subsection{Deep gap regime --- massive planets}
\label{sub:map-deep-gap}

While our radiative model in the gap opening regime (Sect.~\ref{sec:gap-opening-regime}) shows the expected behavior where the planet opens a gap, it lacks a very common feature of simulations with massive, gap-opening planets: a large vortex at the outer gap edge. Such vortices are quite common in models of giant, gap-opening planets embedded in protoplanetary disks \citep[e.g.,][]{hammer-etal-2021,hammer-lin-2023}, and are especially long-lived for low viscosities \citep{rometsch-etal-2021} and isothermal or adiabatic conditions \citep{fung-ono-2021,rometsch-etal-2021}.

Vortices in this context are important for the planet's migration behavior, as they can drive the planet inwards by diffusively refilling the gap in a process termed ``vortex-driven'' migration \citep{lega-etal-2022}. While this is quite similar to the ``vortex-assisted'' migration discussed by \citetalias{MCNALLY-ETAL-2019A} and in Sect.~\ref{sec:feedback-regime}, it is characterized by a single, massive, long-lived vortex rather than a multitude of small-scale, intermittent vortices. For this reason, we distinguish between the gap-opening and deep gap regimes in our map, with the latter representing the presence of a large vortex at the outer gap edge. Nevertheless, we still need to address why such vortices are not seen in our model, and how relevant this ``vortex-driven'' migration is in the long term (i.e., on migration timescales).

The reason why a single large vortex is absent in our radiative model is most likely that the disk scale height, which sets the maximum size of a vortex, is quite small ($h\sim0.025$ at 1\,au). As a result, even the largest vortices remain relatively small, and can be disrupted by radiative diffusion or cooling, leading instead to the wave-like patterns observed in Fig.~\ref{fig:thermal-rad-gap}. Nevertheless, it is possible that a sharper gap edge could result in stronger vortex formation, (e.g., by a more massive planet). In the aforementioned studies featuring long-lived, large-scale vortices, the planet's mass typically exceeds $10\,\Mth$ and the disk's aspect ratio is larger ($h\gtrsim0.05$), placing those models in a different regime of planet--vortex interaction altogether \citep[see, however,][where a sub-thermal mass planet generates a single large vortex for $h\sim0.05$, consistent with our expectations]{dong-etal-2017}.
In addition, in our models, the gap opening process is further assisted by radiative cooling, the effects of which trigger later during the gap opening process. This most likely helps suppress the formation of a large vortex at the outer gap edge, as the gap deepens over a longer timescale \citep[see also][]{hammer-etal-2017}.

Regarding the relevance of vortex-driven migration, \citet{lega-etal-2022} have shown with a suite of simulations with simplified thermodynamics (a constant $\beta\sim1$) that once the large vortex has dissipated, the planet reverts to type-II migration. Given that the disk is not evolving viscously ($\alpha=0$ in their models), the planet stalls, similar to the behavior seen in Sect.~\ref{sec:gap-opening-regime}. This vortex-driven regime is therefore transient, and will merge with the gap opening regime discussed in Sect.~\ref{sub:map-gap-opening} once the vortex has dissipated. This can happen due to several mechanisms such as cooling \citep{fung-ono-2021,rometsch-etal-2021}, the elliptical instability \citep{lesur-papaloizou-2009}, and---to an extent---dust--gas interaction \citep{raettig-etal-2021,lovascio-etal-2022}, which can collectively eliminate the vortex on timescales of $\lesssim10^3$\,years, therefore not affecting planet migration on disk evolution timescales ($\sim10^6$\,years).

In conclusion, while we make the distinction between the gap opening ($\Mp\gtrsim\Mth$) and a deep gap opening regime ($\Mp\gg\,\Mth$) in our map, we stress that the two are not fundamentally different in terms of the planet's migration behavior. Once a deep gap has been established and any associated vortices have dissipated, the planet will stall or migrate inwards at a type-II rate, depending on the disk's viscosity. This allows us to extend the applicability of our results to planets much larger than the thermal mass, as long as the planet's mass is not so large that it can intercept the MHD wind-driven accreting flow through the disk surface \citep[][see also Sect.~\ref{sub:discussion-mhd}]{nelson-etal-2023}.

\subsection{Unrealistic regimes}
\label{sub:map-unrealistic}

In our adiabatic models in Sects.~\ref{sec:feedback-regime}~\&~\ref{sec:gap-opening-regime} we assumed that entropy is conserved during the planet--disk interaction process. This implies that the entropy generation due to the planet's spiral shocks (and therefore shock heating) as well as radiative effects that can relax the local entropy are both negligible. This was a reasonable assumption for the low-mass planet in Sect.~\ref{sec:feedback-regime}, but breaks down when either the conservation of entropy or the assumption of a nearly infinite cooling timescale are violated.

In the context of gap opening planets, as discussed in Sect.~\ref{sub:map-gap-opening}, the existence of a gap renders the disk in the planet's vicinity optically thin, such that cooling effects can become relevant. This results in a significantly different temperature profile within the gap region compared to radiative models (see bottom panel of Fig.~\ref{fig:gaps-thermal}), in addition to missing the critical feedback loop that triggers the gap opening process.

As the planet's mass increases, spiral shocks become more prominent and the assumption that shock heating is negligible can no longer be justified. This is already visible in Fig.~\ref{fig:gaps-thermal}, where the background disk is slightly hotter than the initial equilibrium profile in the radiative model, but becomes significantly more important for more massive planets in similar disk conditions \citep{zhu-etal-2015,rafikov-2016,ziampras-etal-2020a}. In this case, the heat input by the planet via spiral shocks can substantially alter the disk temperature, in turn modifying the torques experienced by the planet, the gap opening process, and even the definition of the thermal mass. However, an adiabatic assumption by definition misses this effect, and the planet's migration behavior is not accurately captured.

In light of the above, super-thermal-mass planets are fundamentally incompatible with the adiabatic assumption, and for this reason we mark the top right corner of the map ($\Mp\gtrsim\Mth$, ``adiabatic'') as unrealistic. In fact, both gap opening as well as shock heating will prompt a radiative response by the disk in order to maintain thermal equilibrium (or even stability in the case of shock heating), such that that section of the map can be absorbed into the ``radiative'' regime, naturally extending the applicability of our results.

\subsection{Planet formation tracks}
\label{sub:map-formation}

Having discussed the different regimes in our map, we can now consider the migration track of a planet as it forms in the disk. In the right panel of Fig.~\ref{fig:butcher-map} we sketch some possible formation tracks for planets forming both interior and exterior to the water snowline, accreting material as they migrate. The background disk model is identical to that in Sect.~\ref{sec:gap-opening-regime}, and the distinction between different regimes is based on Eqs.~\eqref{eq:thermal-mass} and \eqref{eq:cooling-timescale}. The tracks, however, are purely illustrative. Based on the above discussion, we can make several observations.

The shaded regions on the two sides of the map represent the regimes where planet--disk interaction can be well-approximated by an adiabatic (left) or locally isothermal (right) equation of state. These regions, and especially the adiabatic regime, are curved to qualitatively capture the emergence of radiative effects as the gap opening process progresses. As a result, the most impactful takeaway from the map is that radiative processes will be important for the vast majority of planets forming in the disk, even though this might not be immediately obvious from the map on the left panel of Fig.~\ref{fig:butcher-map}. 

Planets forming interior to the water snowline are expected to quickly enter the feedback regime due to the low thermal mass in the inner disk, migrate inwards a short distance, and then stall due to gap opening. Depending on the planet's accretion rate, it is possible for a planet to form practically in situ or for it to migrate closer to the star before stalling. The surviving planets are expected to be rocky.

Planets forming further out spend a substantial amount of time in the linear regime due to their mass initially being small even compared to the feedback mass. As they grow, they eventually transition to the feedback and finally the gap-opening regime, where they stall. Depending on their accretion rate, it is possible to form super-Earths with a nonzero water fraction, mini-Neptunes, or even gas giants.

In both scenarios, for low enough accretion rates, it is possible for the planet to migrate through the entire disk and reach the central star before reaching the gap opening regime. As a result, it is crucial to constrain the planetary accretion process in order to accurately predict the migration track of a forming planet. Nevertheless, our sketch highlights the extent to which radiative cooling can affect the migration behavior of a forming planet, and shows that it is possible to form a wide variety of planets with different orbital properties and compositions within this framework.

\subsection{Limitations and further work}
\label{sub:map-limitations}

We stress that our analysis does not cover the effects of dust--gas interaction or MHD processes, which are discussed in detail in Sects.~\ref{sub:discussion-dust}~\&~\ref{sub:discussion-mhd}. The positive, dust-related corotation torque can be relevant in the type-I regime under certain conditions \citep{benitez-pessah-2018,guilera-etal-2023}, and the presence of a laminar, MHD-driven flow can similarly drive outward migration \citep{mcnally-etal-2017,kimmig-etal-2020}. Furthermore, the luminous feedback due to pebble accretion onto the planet can drive an additional, positive ``thermal torque'' \citep{benitez-etal-2015,cornejo-etal-2023}, which can be relevant for low-mass planets in the type-I regime. As the planet approaches the sub-au regions, turbulence emanating from the inner rim of the disk due to the MRI \citep{flock-etal-2017a,iwasaki-etal-2024} can also help stall migrating planets \citep{chrenko-etal-2022}

Regarding massive planets in the type-II regime, the transition to a deep gap can be accelerated by the presence of a magnetic field \citep{aoyama-bai-2023}, and for planets much larger than the thermal mass it is possible that the planet can even intercept the MHD wind-driven accreting flow through the surface of the disk, resulting in inward migration \citep{lega-etal-2022,nelson-etal-2023}. MHD effects in general can also affect the background disk structure that the planet is embedded in, complicating planet--disk interaction \citep[e.g.,][]{wafflard-fernandez-lesur-2023}.

Overall, while the map presented here is a useful conceptual tool to understand the different migration regimes in nearly inviscid, radiative disks, it is by no means exhaustive. Further work is necessary to address additional physical processes in protoplanetary disks, but for the sake of clarity and to maintain a well-defined scope we chose to omit existing work on such processes during the construction of this map. Nevertheless, our map reveals the shortcomings of the isothermal approximation, with direct implications for population synthesis modeling and the interpretation of observed exoplanet demographics.

\section{Discussion}
\label{sec:discussion}

In this section we discuss the implications of our results for population synthesis models, and highlight the effects of a more realistic disk model. We also address in more detail the role of dust grains and MHD processes in the context of our simulations.

\subsection{Implications for population synthesis modeling}
\label{sub:discussion-popsynth}

Our results highlight the importance of including radiative cooling in planet migration models, as well as accounting for the thermal structure of the background disk. In doing so, they also point out a few important consequences for population synthesis modeling, where the planet--disk interaction process is encapsulated with parametrized migration rates depending on the planet's mass and the background disk properties.

In the feedback regime discussed in Sect.~\ref{sec:feedback-regime}, the fact that the planet continues to drift inwards via vortex-assisted migration provides further proof that the concept of the ``inertial limit'' is not applicable even in the context of radiation hydrodynamics. The presence of additional baroclinic forcing around the horseshoe region in addition to the resolution-dependent nature of the type-III episodes \citepalias{MCNALLY-ETAL-2019A} further complicate the picture, making it difficult to predict the migration track of a super-Earth in this regime. Ideally, a suite of expensive, high resolution simulations would be needed to make an attempt at capturing the behavior of vortices in a parametrized form. While this is beyond the scope of this work, we note that it is a problem that must be addressed as the planet masses in question are of the order of 1--2\,$\Mearth$, well within the mass range modeled with population synthesis and observed in exoplanet demographics.

In the gap-opening regime analyzed in Sect.~\ref{sec:gap-opening-regime}, we showed that while the process of vortex-assisted migration remains present, it is overshadowed by the interplay between gap opening and radiative cooling. The latter accelerates the carving of a deep gap, causing the planet to practically stall. By making the reasonable assumption that this interplay is indeed a feedback loop that will trigger when the cooling timescale in the gap region drops below $\beta\sim100$ (see Fig.~\ref{fig:beta-radiative}), the condition for a planet to stall is that it is massive enough to carve a deep enough gap before migrating through it. This requirement is effectively addressed by the thermal mass $\Mth$ in Eq.~\eqref{eq:thermal-mass}, which can be reasonably well constrained as it is only relies on the background disk temperature.

In our models, we showed that a planet with $\Mp\approx0.44\,\Mth$ falls within the feedback regime, while one with $\Mp\approx2\,\Mth$ stalls due to gap opening. It is reasonable to assume that the transition between these two regimes happens somewhere in between, and this transition point can then be readily used by population synthesis models to tune their migration rates. We intend to address this in followup work.

\subsection{A more sophisticated dust model}
\label{sub:discussion-dust}

The role of dust grains in our models is reduced to carriers of opacity, as we do not include a true dust component in our simulations. Nevertheless, the asymmetric distribution of dust grains in the corotating region has been shown to exert a net torque on low-mass planets, influencing their migration \citep{benitez-pessah-2018,guilera-etal-2023}. According to \citet{guilera-etal-2023}, this effect is noticeable for either large enough pebbles (with a Stokes number $\St\gtrsim0.01$), or for a high enough initial dust-to-gas mass ratio ($\varepsilon\gtrsim0.1$).

In our disk models, assuming that a significant fraction of the dust reservoir is in the form of mm-sized pebbles due to dust growth \citep{birnstiel-2023}, we can estimate the Stokes number as
\begin{equation}
    \label{eq:stokes}
    \St = \frac{\pi}{2}\frac{\rho_\text{d}\,a_\text{d}}{\Sigma},
\end{equation}
with $a_\text{d}=1$\,mm being the grain radius and $\rho_\text{d}\sim1\,\text{g}/\text{cm}^3$ being a typical value for the grain density. Using $\Sigma\approx500\,\text{g}/\text{cm}^2$ at 1\,au, we find that $\St\approx6\times10^{-4}$, which is well below the threshold for the dust-related torque to become important.

Furthermore, in our simulations we use the opacity model of \citet{lin-papaloizou-1985}, which for the temperature range of our disk ($T\lesssim200$\,K) results in the total opacity being dominated by small dust grains, for which we further assume the dust-to-gas mass ratio is constant at $\epsilon=0.01$. Several improvements can be made, as more modern opacity models that take into account observational constraints of the dust size distribution are available nowadays \citep[e.g.,][]{woitke-etal-2016,birnstiel-etal-2018}. Dust growth will also affect the dust-to-gas mass ratio in small grains \citep{birnstiel-2023}.

At the same time, however, we note that our models are placed at the inner disk regions of $R\lesssim1$\,au, where the disk is expected to be optically thick. This means that even for our models in Sect.~\ref{sec:gap-opening-regime} with $\beta\sim3000$ before gap opening, the disk would remain optically thick even if a more sophisticated dust model was used. This would be the case as well for our models in Sect.~\ref{sec:feedback-regime}, with $\beta\sim10^4$.

Overall, while an accurate model of both the dust--gas dynamics and the dust opacity is important for understanding the migration of super-Earth-mass planets, we expect that neither effect would significantly alter the overall behavior of the planet in our models. We nevertheless highlight that both effects are crucial for the formation of exoplanetary atmospheres \citep[e.g.,][]{szulagyi-etal-2022,krapp-etal-2024}, the flow of dust grains around a planet \citep[e.g.,][]{binkert-etal-2023}, or planet-driven substructures \citep[e.g.,][]{ziampras-etal-2025}.

\subsection{Magnetohydrodynamical effects}
\label{sub:discussion-mhd}
\begin{figure}
    \includegraphics[width=\columnwidth]{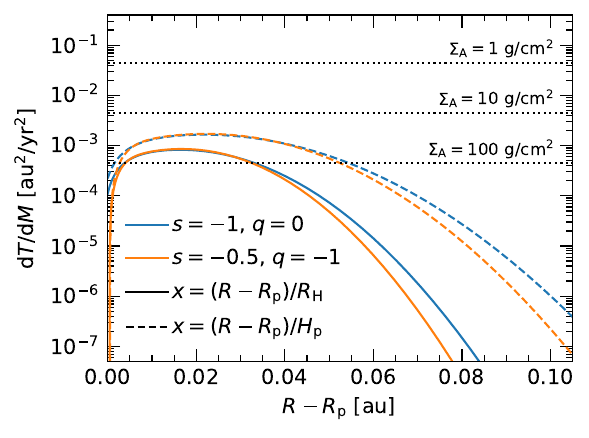}
    \caption{Comparison between planet (blue and orange curves) and wind-driven torques (black dots), showing that the flow will be unimpeded by the planet for $\Sigma_{\rm A} \lesssim 10\,\text{g}/\text{cm}^2$ and blocked by the planet for $\Sigma_{\rm A} \gtrsim 100\,\text{g}/\text{cm}^2$. Solid and dashed lines correspond to gap-opening and low-mass planets, respectively. Different colors denote combinations of $s$ and $q$ following \citet{dangelolubow2010}, showing the spread of the estimated torque.}
    \label{fig:torque-comparison}
\end{figure}

Although we have considered non-accreting disks with very low viscosities, protoplanetary disks are observed to accrete onto their central stars at a canonical rate of ${\dot M} \sim 10^{-8}\,\Msun/\text{yr}$ \citep{hartmann-etal-1998}. According to recent models, accretion occurs via laminar radial gas flows, driven by large-scale magnetic fields, with the details depending on the orientation of the magnetic field among other factors. In regions just interior to 1\,au, and in the presence of a vertical magnetic field, a magnetized wind is launched from the disk surface and accretion occurs in narrow regions near the surface \citep{bai2013,gressel2015,lesur2023}. If the vertical magnetic field vector is aligned with the disk angular momentum vector, then the Hall shear instability can generate large-scale horizontal fields near the midplane, inducing accretion there as well \citep{lesur2014}. Hence, a magnetized disk can sustain laminar gas flows either in narrow layers near the surface or throughout most of its vertical column.

\cite{lega-etal-2022} and \cite{nelson-etal-2023} considered migration and gas accretion by a Jovian mass planet embedded at 5\,au in a simple model of a magnetized disk, which contains narrow accretion layers near each disk surface, overlying an extensive dead zone and providing a total radial mass flux of ${\dot M} = 10^{-8}\,\Msun/\text{yr}$. They used 3D hydrodynamical simulations with an external torque prescription designed to mimic the effects of a magnetized wind. They showed that the migration and accretion rates depend critically on whether the tidal torque due to the planet can block the accretion flow.

The presence of a dead zone guarantees that a planet exceeding the thermal mass will open a deep gap. \cite{lega-etal-2022} showed that when the accretion flow from the outer disk towards the planet is blocked by its tidal torque, the planet can migrate inwards quickly on less than Myr timescales. This is because gas builds up outside the planet's orbit and increases the magnitude of the (negative) torque from this region. As the planet moves inwards the accretion flow fills in the gap behind the planet and sustains its migration. If the accretion flow is unimpeded, however, migration is very slow because the low column density of the accreting layer adds only a small amount of gas in the gap and hence barely changes the migration torque.

The question of whether or not an accretion flow through a disk ought to be blocked by a planet is clearly relevant for our current study. While we cannot include an accretion flow in our vertically averaged 2D simulations, we can estimate whether it would be blocked by the planet.

Using the approach described in \cite{nelson-etal-2023}, we can compare the torque per unit mass acting on the disk due to a planet with a mass of $6.7\,\Mearth$ to the torque required to drive an accretion flow with total mass flux ${\dot M} = 10^{-8}\,\Msun/\text{yr}$, where this is assumed to arise because identical accretion layers are present in each disk hemisphere. 
The torque per unit mass driving accretion in a layer located near one of the disk's surfaces is given by 
\begin{equation}
    \Gamma_{\rm wind} = \frac{\dot M}{8 \pi \Sigma_{\rm A}} \sqrt{\frac{\G M_{\star}}{\Rp^3}},
    \label{eqn:WindTorque}
\end{equation}
where $\Sigma_{\rm A}$ is the column density of the accreting layer, and the expression is to be evaluated at the planet's orbital location $\Rp$.
From \cite{dangelolubow2010}, the torque per unit mass from the planet is given by 
\begin{equation}
    \Lambda_{\rm p}= {\cal F}(x,s,q) q_{\rm p}^2 \Rp^2 \Omega_{\rm p}^2 \hp^{-4},
    \label{eqn:TorquePlanet}
\end{equation}
where $q_{\rm p}=\Mp/\Mstar$ and ${\cal F}(x,s,q)$ is a function that depends on the power-law indices for the surface density and temperature profiles (see Eq.~\eqref{eq:initial-conditions}) and the normalized distance from the planet $x=(R-\Rp)/\Hp$ or $(R-\Rp)/\Rh$ for low-mass and gap-opening planets, respectively, with $\Rh=\Rp\sqrt[3]{q_\text{p}/3}$ being the planet's Hill radius. For a definition of $\mathcal{F}$, we refer the reader to \citet{dangelolubow2010}.

Figure~\ref{fig:torque-comparison} shows a comparison between these torques for different values of $\Sigma_{\rm A}$, assuming $\Rp=1$\,au. For $\Sigma_{\rm A} \le10\,\text{g}/\text{cm}^2$, we see that the accretion flow can pass through the gap and past the planet unimpeded because $\Gamma_{\rm wind} > \Lambda_{\rm p}$. Hence, a 6.7\,$\Mearth$ super-Earth embedded in a radiative disk with surface accretion flows having $\Sigma_{\rm A} \le 10\,\text{g}/\text{cm}^2$ should display very slow or stalled migration as shown in Sect.~\ref{sec:gap-opening-regime}.
However, if the column density of the accreting layer approaches $\Sigma_{\rm A}=100\,\text{g}/\text{cm}^2$, then the flow will start to be blocked and faster migration is likely to ensue. This situation would apply if the Hall effect induces an accretion flow with ${\dot M} = 10^{-8}\,\Msun/\text{yr}$ throughout the vertical column of the disk \citep[see also][]{mcnally-etal-2018,kimmig-etal-2020}.

For the lower mass planets in the feedback regime we have considered, the situation is more complicated even when considering a highly simplified magnetized disk model. The migration behavior in this case depends on the structure within the gap, and this would most likely be modified by magnetic torques, perhaps changing the emergence and influence of vortices.

Recent non-ideal MHD simulations applied to giant planets embedded in the outer regions of protoplanetary disks show that indeed magnetic forces can dramatically change the structure of the gap and influence the migration torques \citep{aoyama-bai-2023,wafflard-fernandez-lesur-2023}. Hence, an important next step is to perform an MHD study of the interaction between a magnetized disk and an embedded super-Earth for conditions that pertain to regions interior to 1\,au. However, the requirement to perform 3D MHD simulations with sufficient resolution to capture the formation and evolution of small-scale vortices presents a formidable computational challenge.

\subsection{Implications for pebble accretion}

From Fig.~\ref{fig:lowmass-adb-gap-structure}, it becomes clear that even a $1.5\,\Mearth$ mass planet can form a pressure bump exterior to its orbit for low enough viscosities. This can be problematic for the planet's growth through pebble accretion \citep[see][]{ormel-klahr-2010,ormel-2024}, as the pressure bump can effectively shut off the pebble flux onto the planet's orbit.

Nevertheless, several pathways might exist to circumvent this issue. In the inner disk, mm-sized pebbles have a Stokes number of $\sim10^{-4}$--$10^{-3}$ (see Eq.~\eqref{eq:stokes}) and would therefore be well-coupled to the gas. As a result, these particles can simply leak through the pressure bump and reach the planet. For larger Stokes numbers (larger pebbles or further out in the disk), the presence of vortices can also maintain a local turbulent diffusivity that can transport pebbles across the gap and near the planet's orbit \citep[see e.g.,][]{cummins-etal-2022}. An alternative scenario is that the planet could have formed further out in the disk, where it would migrate at a type-I rate until it reached the feedback mass, by which point it would already have grown to a typical pebble isolation mass of $\sim5$--$20\,\Mearth$ \citep{lambrechts-etal-2014,bitsch-etal-2018}.

\section{Summary}
\label{sec:summary}

We have presented a set of 2D high-resolution radiative hydrodynamical simulations of planet--disk interaction in nearly inviscid disks. Our focus was the migration behavior of super-Earth-mass planets and in particular the role of radiative cooling in the feedback and gap-opening regimes.

We first explored planet migration in the feedback regime, which translates to $\Mp\gtrsim1\,\Mearth$ for our disk model at 1\,au. Here, we broadly recover the results of \citetalias{MCNALLY-ETAL-2019A} in that the planet migrates inwards in a vortex-assisted, turbulent type-I-like fashion, in contrast to the proposed ``inertial limit'' which would cause the planet to stall. Radiative effects delay the onset of the Rossby wave instability, reducing the frequency of potential runaway migration episodes, but also induce baroclinic forcing in the horseshoe region that encourages an inward drift. While the combined effects ultimately lead to sustained inward migration, their individual contributions can have implications for the observation of planet-driven features (e.g., substructures due to the pressure traps formed after a type-III episode), as the planet is expected to migrate overall faster but with less frequent type-III episodes.

For gap-opening planets ($\Mp\gtrsim4\,\Mearth$ at 1 au), we identified a new regime where the planet stalls at the center of a deep gap rather than migrating inwards in a vortex-assisted fashion. This mechanism is the result of the interplay between the gap opening process and radiative cooling, which accelerates the carving of a deep gap and causes the planet to transition to a type-II-like migration regime, which is much slower than the disk lifetime. This effect cannot be captured by either adiabatic or isothermal models---or even by models with a fixed cooling timescale---as it relies on the gap region becoming marginally optically thin ($\beta\sim1$) dynamically \citep[see also][for an equivalent situation for circumbinary disks]{sudarshan-etal-2022}. Given that the median mass of super-Earth-mass planets is $5.6\,\Mearth$, this regime is expected to be of central relevance for planet population synthesis models, as it provides a natural stalling mechanism for planets that would otherwise migrate inwards.

By combining our results in the feedback and gap-opening regimes---with or without radiative effects---with previous studies in the type-I regime as well as for very massive planets, we constructed a map of planet migration regimes in nearly inviscid, radiative disks. This map, inspired by the work of \citetalias{MCNALLY-ETAL-2019A}, provides a conceptual framework for understanding the migration behavior of super-Earth-mass planets in protoplanetary disks, and highlights the importance of radiative cooling in determining the planet's migration track.

In this study we have achieved two main goals. With dedicated models, we have shown that radiative effects can significantly affect the migration track of the ``typical super-Earth'', which finds applications in both population synthesis as well as planet--disk interaction in general. By then aggregating previous related work, we have developed a framework of planet migration in nearly inviscid disks that can be extended to include additional physical processes, such as dust--gas interaction or MHD effects. While several sections of the map are either underexplored or subject to additional physics, we believe that it provides a useful milestone for future work on planet migration.

\section*{Acknowledgments}
We thank the referee for their constructive feedback, which helped improve the clarity of this work.
AZ would like to thank Til Birnstiel, Remo Burn, Thomas Henning, and Martin Pessah for their helpful advice in connecting this work to population synthesis models and for their constructive feedback. This research utilized Queen Mary's Apocrita HPC facility, supported by QMUL Research-IT (http://doi.org/10.5281/zenodo.438045). This work was performed using the DiRAC Data Intensive service at Leicester, operated by the University of Leicester IT Services, which forms part of the STFC DiRAC HPC Facility (www.dirac.ac.uk). The equipment was funded by BEIS capital funding via STFC capital grants ST/K000373/1 and ST/R002363/1 and STFC DiRAC Operations grant ST/R001014/1. DiRAC is part of the National e-Infrastructure. AZ and RPN acknowledge support from the STFC grants ST/X000931/1 and ST/T000341/1. AZ acknowledges funding from the European Union under the European Union's Horizon Europe Research and Innovation Programme 101124282 (EARLYBIRD). Views and opinions expressed are those of the authors only. All plots in this paper were made with the Python library \texttt{matplotlib} \citep{hunter-2007}. Typesetting was expedited with the use of GitHub Copilot, but without the use of AI-generated text.

\section*{Data Availability}

Data from our numerical models are available upon reasonable request to the corresponding author.

\bibliographystyle{mnras}
\bibliography{refs}


\appendix

\section{Resolution check}
\label{apdx:resolution}

As mentioned in Sect.~\ref{sec:comparison-to-mcnally}, we chose an $\alpha=10^{-6}$ in order to probe the nearly inviscid regime without exposing our model to numerical diffusion \citepalias[see][for a detailed discussion]{MCNALLY-ETAL-2019A}. It is nevertheless useful to verify that our results are not qualitatively resolution-dependent, especially in the feedback regime, where the planet's migration behavior is determined by the formation of small-scale vortices.

\begin{figure}
    \includegraphics[width=\columnwidth]{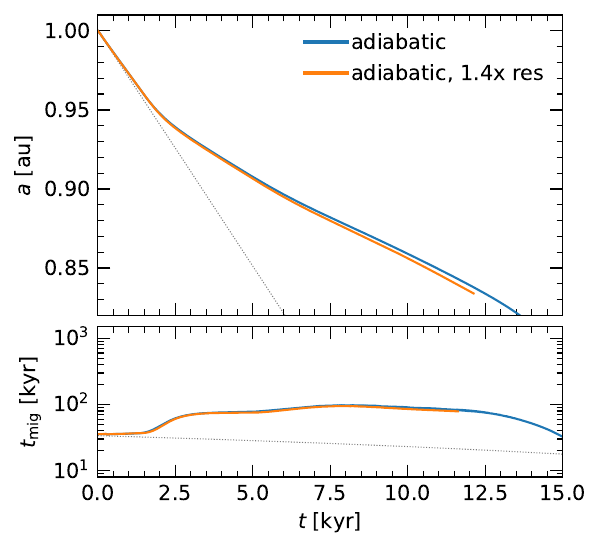}
    \caption{Planet tracks for our fiducial adiabatic model from Sect.~\ref{sec:feedback-regime} (blue) and a higher resolution model (orange). The two show excellent agreement in terms of both migration tracks and timescales.}
    \label{fig:resolution}
\end{figure}

To that end, we repeated our adiabatic model in Sect.~\ref{sec:feedback-regime} with a $1.4\times$ higher grid resolution for $N_R\times N_\phi=2688\times9072$ until the planet entered the vortex-assisted migration phase, evident by a flattening of the migration timescale. The results are shown in Fig.~\ref{fig:resolution}, where we compare the migration tracks and timescales of the two models. We find an excellent agreement between the two regarding the exact time that the planet transitions from a type-I behavior to the feedback regime, as well as the onset of the vortex-assisted phase. We can therefore conclude that our models are robust against numerical dissipation, especially considering that our radiative models feature additional, radiative diffusion of physical origin.

\bsp	
\label{lastpage}
\end{document}